\documentclass[%
reprint,
superscriptaddress,
showpacs,
nofootinbib,
amsmath,amssymb,
prc,
floatfix]%
{revtex4-1}

\usepackage[pdftex]{graphicx}
\usepackage{bm}

\usepackage{braket}

\allowdisplaybreaks

\begin{document}

\title{
Coordinate-space solver for finite-temperature Hartree-Fock-Bogoliubov
calculation using the shifted Krylov method
}

\author{Yu Kashiwaba}%
\email{kashiwaba@nucl.ph.tsukuba.ac.jp}
\affiliation{Faculty of Pure and Applied Sciences,
              University of Tsukuba, Tsukuba 305-8577, Japan}

\author{Takashi Nakatsukasa}%
\email{nakatsukasa@nucl.ph.tsukuba.ac.jp}
\affiliation{Center for Computational Sciences,
              University of Tsukuba, Tsukuba 305-8577, Japan}
\affiliation{Faculty of Pure and Applied Sciences,
              University of Tsukuba, Tsukuba 305-8577, Japan}
\affiliation{RIKEN Nishina Center, Wako 351-0198, Japan}

\date{\today}

\begin{abstract}
	\noindent
{\bf Background:}
In order to study structure of proto-neutron stars and those in
subsequent cooling stages,
it is of great interest to calculate inhomogeneous hot and cold nuclear matter
in a variety of phases.
The finite-temperature Hartree-Fock-Bogoliubov (FT-HFB) theory is
a primary choice for this purpose, however, its numerical calculation for
superfluid (superconducting) many-fermion systems in three dimensions requires
enormous computational costs.
\\
{\bf Purpose:}
To study a variety of phases in the crust of hot and cold neutron stars,
we propose an efficient method to perform the FT-HFB calculation
with the three-dimensional (3D) coordinate-space representation.
\\
{\bf Methods:}
Recently, an efficient method based on the contour integral of Green's function with
the shifted conjugate-orthogonal conjugate-gradient method has been
proposed [Phys. Rev. C {\bf 95}, 044302 (2017)].
We extend the method to the finite temperature,
using the shifted conjugate-orthogonal conjugate-residual method.
\\
{\bf Results:}
We benchmark the 3D coordinate-space solver of the FT-HFB calculation
for hot isolated nuclei and fcc phase in the inner crust of neutron stars
at finite temperature.
The computational performance of the present method is demonstrated.
Different critical temperatures of the
quadrupole and the octupole deformations are confirmed for $^{146}$Ba.
The robustness of the shape coexistence feature in $^{184}$Hg
is examined.
For the neutron-star crust,
the deformed neutron-rich Se nuclei embedded in the sea of superfluid
low-density neutrons
appear in the fcc phase
at the nucleon density of 0.045 fm$^{-3}$ and the temperature of
$k_B T=200$ keV.
\\
{\bf Conclusions:}
The efficiency of the developed solver is demonstrated for nuclei
and inhomogeneous nuclear matter at finite temperature.
It may provide a standard tool for nuclear physics, especially for
the structure of the hot and cold neutron-star matters.
\end{abstract}

\maketitle

\section{\label{sec:intro}Introduction}

The mean-field approaches, such as Hartree-Fock (HF)
and Hartree-Fock-Bogoliubov (HFB) theories, 
have been playing a central role in studying heavy nuclei
and nuclear matter \cite{BHR03}.
They are especially useful for studies of the ground (stationary) states.
In addition, 
the time-dependent extension of the mean-field theories is straightforward
and provides a powerful tool for studies of nuclear response and reaction
\cite{Nak12,Sim12,Bul13,NMMY16}.
Including the pairing correlations, a number of calculations have
been performed with the BCS approximation \cite{Eba10,ENI14,SL14,SSL15}.
Recently, studies of three-dimensional (3D) nuclear dynamics using
the full time-dependent Hartree-Fock-Bogoliubov (TDHFB) method
have become available
\cite{SBMR11,Has12,Has13,SBBMR15,HS16,MSW17,SH17,SH19}.
The time evolution of the TDHFB states requires calculations of
all the time-dependent quasiparticle states, which is computationally
very demanding.

The static HFB calculation 
seems to be easier than the time-dependent problems,
at first sight.
However, in fact, it is often more difficult than the the time-dependent
calculation.
This is due to requirement of the self-consistency between the HFB state
and the HFB Hamiltonian.
A standard procedure is as follows.
The diagonalization of the HFB Hamiltonian produces the quasiparticle
states.
The quasiparticle states define the normal and pair densities
which determine the HFB Hamiltonian.
The iteration is necessary to reach the self-consistency.
For the full 3D unrestricted calculations, finding a self-consistent solution
is not as simple as it might seem.
It involves successive diagonalization of matrices with large dimension $N$,
which normally needs operations of $O(N^3)$.
Most of available codes of the HFB calculation
utilize some symmetry restriction on the
densities, such as spatial symmetry and time-reversal symmetry,
in order to reduce both the matrix dimension and the number of iteration
\cite{HFBRAD,HFBTHO17,Ev8}.
The HFB program {\sc hfodd} \cite{HFODD17} is able to perform the unrestricted
calculation, however, since it is based on the harmonic-oscillator basis,
it is difficult to calculate nuclei near the neutron drip line
and various phases of nuclear matter in the neutron stars.

Recently, a novel computational approach to the HFB iterative problem
has been proposed by Jin, Bulgac, Roche, and Wlaz{\l}owski
\cite{JBRW17}.
In contrast to the conventional methods,
this approach has several favorable aspects,
especially in large-scale calculations.
(1) The densities are calculated by the contour integral
in the complex energy plane, without quasiparticle wave functions.
The matrix diagonalization is unnecessary.
(2) It is based on the shifted Krylov subspace method for calculating
the Green's function $G(\bm{r},\bm{r}';z)$ with complex energies $z$. 
The shifted method allows us to obtain simultaneously the Green's
function with different energies $z$.
(3) It is suited for parallel computing, because the 
Green's function $G(\bm{r},\bm{r}';z)$ is calculable
independently for each point $\bm{r}'$.
In Ref.~\cite{JBRW17}, they explored the shifted conjugate-orthogonal
conjugate-gradient (COCG) method, and showed a few benchmark calculations
with the coordinate-space representation.

In this paper, we propose an extension of the 3D coordinate-space
HFB method of Ref.~\cite{JBRW17} to that at finite temperature,
namely the finite-temperature HFB (FT-HFB) calculation.
The FT-HFB method is valuable for studying a variety of aspects in
nuclear and many-fermion systems.
For example, the structure and the composition of
the (proto-)neutron stars depend on
the equation of state (EOS) of baryonic matter at finite temperature.
In order to calculate inhomogeneous baryonic matter in the crust region,
the 3D coordinate-space FT-HFB solver is highly desired.
For experimental studies on nuclear structure,
giant dipole resonances in hot nuclei provide us
information on nuclear shapes at finite temperature \cite{ST07}.
In order to study shape change together with pairing and shell quenching
in hot nuclei, the FT-HFB is a valuable tool.
The shape dynamics at finite temperature may play an important role
in induced fission processes \cite{PNSK09,SNP09}.
The FT-HFB has been also utilized to study
the level density \cite{HGGK12,FMI19}, 
which is one of the key ingredients in
the statistical reaction model.

The paper is organized as follows.
In Sec.~\ref{sec:formulation}, we recapitulate the FT-HFB theory,
then, present a computational method of the contour integral
to produce the normal and abnormal densities.
It is slightly more complicated than the zero-temperature HFB
in Ref.~\cite{JBRW17},
because we need to remove contributions from the Matsubara frequencies on the imaginary axis.
In Sec.~\ref{sec:results}, we demonstrate some numerical results.
Finally, the summary and the perspectives are given in Sec.~\ref{sec:summary}.

\section{\label{sec:formulation} Theoretical formulation}

In this section, we first recapitulate the FT-HFB theory, then,
introduce the Green's function method for that.
Readers are referred to Ref.~\cite{JBRW17}
for the zero-temperature formulation.

\subsection{Finite temperature HFB theory}

Considering a system of spin 1/2 particles
with the volume $V$ and the Hamiltonian $\hat{H}$,
in a thermal equilibrium with a heat bath of the temperature $T$ and
the chemical potential $\mu$.
The grand partition function is given by
$Z(T,V,\mu)\equiv \mathrm{Tr} \left[ e^{-\beta (\hat{H}-\mu\hat{N}}) \right]$,
where $\beta\equiv (k_B T)^{-1}$, 
and 
$\hat{N}\equiv \sum_\sigma \int_V \hat{\psi}^\dagger(\bm{r}\sigma)
\hat{\psi}(\bm{r}\sigma) d\bm{r}$ is the particle number operator.
In nuclear physics, we need to treat both protons and neutrons (isospin degrees of freedom).
This extension can be easily done by incorporating both proton and
neutron densities when we calculate potentials in Eq.~(\ref{eq:handDelta}).

The mean-field approximation replaces $\hat{H}-\mu \hat{N}$ by
the HFB Hamiltonian which is given in terms of
independent quasiparticles.
Using the quasiparticle number operator
$\hat{n}_k\equiv\hat{\gamma}_k^\dagger \hat{\gamma}_k$ with
the creation and annihilation operators
$(\hat{\gamma}_k^\dagger, \hat{\gamma}_k)$,
The HFB Hamiltonian is simply written as
\begin{equation}
\hat{H}_\mathrm{HFB} =
	E_0 -\mu N_0 +\sum_{k>0} E_k \hat{n}_k ,
	\label{eq:HFB_Hamiltonian}
\end{equation}
where $E_0$ is the energy of the HFB ground state $\ket{0}$
with the particle number $N_0=\bra{0}\hat{N}\ket{0}$,
and $k>0$ means the quasiparticle states with positive energies
$E_k>0$.
The state $\ket{0}$ is defined as the quasiparticle vacuum.
\begin{equation}
\hat{\gamma}_k\ket{0} = 0  \mbox{ for } k>0.
\end{equation}
The trace in the partition function is calculated by summing up
expectation values with respect to all the $n$-quasiparticle states
with $n=0,1,\cdots$.
\begin{equation}
Z_\mathrm{HFB}(T,V,\mu)=e^{-\beta (E_0-\mu N_0)}
	\prod_{k>0} \left( 1+e^{-\beta E_k} \right) ,
\end{equation}
which leads to the density matrix,
\begin{equation}
\hat{\rho}_\mathrm{HFB}(T,V,\mu) =\frac{e^{-\beta \hat{H}_\mathrm{HFB} }}
	{Z_\mathrm{HFB}(T,V,\mu)} 
	=
	\frac{\prod_{k>0}e^{-\beta E_k \hat{n}_k}}
	{\prod_{k>0} \left( 1+e^{-\beta E_k} \right) } .
\end{equation}
Thus, the one-body densities are given as
\begin{eqnarray}
	\label{eq:rho_T}
&&\rho_T(\xi,\xi')
	\equiv\mathrm{Tr} \left[ 
	\hat{\rho}_\mathrm{HFB}
	\hat{\psi}^\dagger(\xi')\hat{\psi}(\xi)
	\right] \nonumber\\
	&&= \sum_{k>0} \left\{
	f_k u_k(\xi) u_k^*(\xi')
	+ (1-f_k) v_k^*(\xi) v_k(\xi')
	\right\} , \\
	\label{eq:kappa_T}
&&\kappa_T(\xi,\xi')
	\equiv\mathrm{Tr} \left[ 
	\hat{\rho}_\mathrm{HFB}
	\hat{\psi}(\xi')\hat{\psi}(\xi)
	\right] \nonumber\\
	&&= \sum_{k>0} \left\{
		(1-f_k) v_k^*(\xi) u_k(\xi')
	+f_k u_k(\xi) v_k^* (\xi')
	\right\} ,
\end{eqnarray}
where $\xi$ indicates the coordinate and spin, $\xi=(\bm{r},\sigma)$,
and the quasiparticle occupation is given by
\begin{equation}
f_k\equiv\frac{1}{e^{\beta E_k}+1} .
\label{eq:occ}
\end{equation}
For Eqs.~(\ref{eq:rho_T}) and (\ref{eq:kappa_T}),
we use the Bogoliubov transformation,
\begin{flalign}
\hat{\psi}^{\dagger} (\xi) &= \sum_{k>0} \left[ u_k^* (\xi) \hat{\gamma}_k^{\dagger} + v_k(\xi)\hat{\gamma}_k \right], 
	\label{eq:BTa}\\
\hat{\psi} (\xi) &= \sum_{k>0} \left[  u_k(\xi)\hat{\gamma}_k + v_k^* (\xi) \hat{\gamma}_k^{\dagger} \right].
	\label{eq:BTb}
\end{flalign}
Using the matrix notation of
\begin{equation}
	U_{\xi k} = u_k(\xi), \quad\quad
	V_{\xi k} = v_k(\xi), \quad\quad
	f_{kk'} = f_k \delta_{kk'} ,
\end{equation}
Eqs.~(\ref{eq:rho_T}) and (\ref{eq:kappa_T}) can be denoted in a compact form.
\begin{equation}
	\rho_T=UfU^\dagger +V^*(1-f)V^T ,\quad
	\kappa_T=UfV^\dagger +V^*(1-f)U^T .
	\label{eq:rho-kappa}
\end{equation}

The quasiparticle energies and wave functions are obtained by
solving the HFB equation
\begin{equation}
\left[
    \begin{array}{cc}
    h & \Delta \\
    -\Delta^* & -h^*
    \end{array}
  \right]
 \left[
    \begin{array}{c}
        u_k  \\
        v_k 
    \end{array}
  \right]
= E_k
 \left[
    \begin{array}{c}
        u_k \\
        v_k 
    \end{array}
  \right],
\label{eq:HFB}
\end{equation}
where $h(\xi,\xi')$ and $\Delta(\xi,\xi')$ are formally given by
the derivatives of the energy density functional (EDF)
$\mathcal{E}[\rho,\kappa]$,
\begin{equation}
	h(\xi,\xi')=\frac{\delta \mathcal{E}}{\delta\rho(\xi',\xi)}
	-\mu\delta(\xi,\xi'),
	\quad
	\Delta(\xi,\xi')=\frac{\delta \mathcal{E}}{\delta\kappa^*(\xi,\xi')} .
	\label{eq:handDelta}
\end{equation}
Here, we use a simplified notation
$\delta(\xi,\xi')\equiv\delta(\mathbf{r}-\mathbf{r}')\delta_{\sigma\sigma'}$.

\subsection{HFB Hamiltonian for Skyrme EDF}
In the Skyrme functional, the nuclear energy is written as
$
	E[\rho,\kappa]=\int d\bm{r} \mathcal{E}(\bm{r}) ,
$
where the energy density is given by the sum of kinetic, nuclear potential,
Coulomb, and pairing energies.
\begin{flalign}
	\mathcal{E} = \mathcal{E}_{\mathrm{kin}} + \mathcal{E}_{\mathrm{nuclear}} + \mathcal{E}_{\mathrm{Coul}} + \mathcal{E}_{\mathrm{pair}} .
\end{flalign}
The energy density is a functional of local densities, such as
normal density
$\rho_q(\bm{r})$, 
kinetic density $\tau_q(\bm{r})$,
spin-current density $\bm{J}_q(\bm{r})$,
and the pair (abnormal) density $\nu_q(\bm{r})$,
with $q=n$ and $p$.
These densities are calculated from the one-body densities,
Eqs. (\ref{eq:rho_T}) and (\ref{eq:kappa_T}).
The energy density of the Coulomb exchange is given by the Slater approximation,
\begin{flalign}
\mathcal{E}_{\mathrm{Coul}} (\bm{r}) = 
	\frac{e^2}{2} \int \frac{\rho_p (\bm{r})\rho_p (\bm{r'})}
	{\left| \bm{r} - \bm{r}' \right|} \mathrm{d}\bm{r}' 
	- \frac{3e^2}{4} \left( \frac{3}{\pi} \right)^{1/3}
	\rho_p^{4/3} (\bm{r}).
\end{flalign}
The pairing energy density depends on the local pairing density,
\begin{flalign}
\mathcal{E}_{\mathrm{pair}}(\bm{r})
	= \sum_{q=n,p} g_{\mathrm{eff}}(\bm{r})
	\left| \nu_q (\bm{r}) \right|^2,
	\label{eq:pairing_energy}
\end{flalign}
where the effective pairing strength $g_{\mathrm{eff}}$
is determined via a renormalization \cite{BY02}
of the bare pairing strength $g_0$.
The adopted value for $g_0$ in the present manuscript is
given in Sec.~\ref{sec:results}.
The local nature of the Skyrme energy density leads to 
the HFB equation (\ref{eq:HFB}) in the coordinate representation of the form
\begin{flalign}
 H_{\mathrm{HFB}} \left(
    \begin{array}{c}
    u_{k\uparrow} \\
    u_{k\downarrow} \\
    v_{k\uparrow} \\
    v_{k\downarrow}
    \end{array}
  \right)= E_k \left(
\begin{array}{c}
    u_{k\uparrow} \\
    u_{k\downarrow} \\
    v_{k\uparrow} \\
    v_{k\downarrow}
    \end{array}
  \right),
	\label{eq:HFB_2}
\end{flalign}
\begin{flalign}
 H_{\mathrm{HFB}}=\left(
    \begin{array}{cccc}
    h_{\uparrow\uparrow} & h_{\uparrow\downarrow} & 0 & \Delta \\
    h_{\downarrow\uparrow} & h_{\downarrow\downarrow} & -\Delta & 0 \\
    0 & -\Delta^* & -h^*_{\uparrow\uparrow} & -h^*_{\uparrow\downarrow} \\
    \Delta^* & 0 & -h^*_{\downarrow\uparrow} & -h^*_{\downarrow\downarrow}
    \end{array}
  \right).
\end{flalign}
Here and hereafter in this section, the isospin index $q=n,p$ is omitted for simplicity.
$h_{\sigma\sigma'}\equiv h(\xi,\xi')$ of Eq. (\ref{eq:handDelta})
with $\xi=(\bm{r}\sigma)$ and
$\xi'=(\bm{r}'\sigma')$,
which are diagonal in the coordinate
except for the derivative terms.
$\Delta$ is strictly diagonal,
$\Delta(\bm{r},\bm{r}')=
\Delta(\bm{r})\delta(\bm{r}-\bm{r}')$ with
\begin{equation}
\Delta(\bm{r})= g_{\mathrm{eff}}(\bm{r})
	\nu (\bm{r}) .
\end{equation}


All the local densities are calculable from the one-body densities,
(\ref{eq:rho_T}) and (\ref{eq:kappa_T}),
at temperature $T$.
\begin{eqnarray}
\rho(\bm{r}) &=& \sum_\sigma \rho_T(\bm{r}\sigma,\bm{r}\sigma) ,
\label{eq:rho}\\
\nu(\bm{r}) &=& \kappa_T(\bm{r}\uparrow,\bm{r}\downarrow) ,
\label{eq:nu}\\
\tau(\bm{r}) &=& \sum_\sigma
	\nabla_1\cdot\nabla_2 \rho_T(\bm{r}\sigma,\bm{r}\sigma) ,
\label{eq:tau}
	\\
\bm{J}(\bm{r}) &=& \frac{1}{2i}
	(\nabla_1-\nabla_2) \times
	\bm{s}(\bm{r},\bm{r}) 
\label{eq:J}
\end{eqnarray}
where $\nabla_{1(2)}$ indicates the differentiation on the first (second) argument
$\bm{r}$ of the densities.
Here, the spin density $\bm{s}(\bm{r},\bm{r}')$
is defined in terms of the Pauli matrix
$\boldsymbol\sigma$ as,
\begin{eqnarray}
\bm{s}(\bm{r},\bm{r}') &=& \sum_{\sigma\sigma'}
	\rho_T(\bm{r}\sigma,\bm{r}'\sigma') \bra{\sigma'}
	\boldsymbol{\sigma} \ket{\sigma} .
\end{eqnarray}

In the present paper, we assume the time-reversal symmetry.
We use the following relations to reduce the computational cost:
\begin{eqnarray}
\rho_T(\bm{r}\sigma,\bm{r}'\sigma')
	=s_\sigma s_{\sigma'}
	\rho_T^*(\bm{r}\bar{\sigma},\bm{r}'\bar{\sigma}') ,
\label{eq:TR1} \\
\kappa_T(\bm{r}\sigma,\bm{r}'\sigma')
	=s_\sigma s_{\sigma'}
	\kappa_T^*(\bm{r}\bar{\sigma},\bm{r}'\bar{\sigma}') ,
\label{eq:TR2}
\end{eqnarray}
where $\bar{\sigma}=(\downarrow,\uparrow)$ for
$\sigma=(\uparrow,\downarrow)$,
and $s_\uparrow=-s_\downarrow=1$.
Thus, we need to calculate only those with $\sigma'=\uparrow$.
All the time-odd densities vanish.

\subsection{Green's functions and local densities}
Now, let us present a method using the Green's function to calculate
the local densities at finite temperature.
The Green's functions of the HFB equation (\ref{eq:HFB_2}),
\begin{flalign}
G(z) = \left(
    \begin{array}{cc}
      G_{uu}(z;\xi,\xi')  & G_{uv}(z;\xi,\xi') \\
      G_{vu}(z;\xi,\xi') & G_{vv}(z;\xi,\xi')  
    \end{array}
  \right) ,
\label{eq:green2x2}
\end{flalign}
are defined by a solution of
\begin{flalign}
\label{eq:shift}
	\left(zI -H_\mathrm{HFB}\right) G(z) = I ,
\end{flalign}
with a proper boundary condition.
Here, $I$ is the unit matrix.
Each element of Eq. (\ref{eq:green2x2}) can be expressed as 
\begin{flalign}
G_{uu}(z;\xi;\xi')
=\sum_{k>0} \left[\frac{u_k (\xi) u_k^{*}(\xi')}{z-E_k}
	+\frac{v^*_k (\xi) v_k(\xi')}{z+E_k} \right], \notag\\
G_{uv}(z;\xi,\xi')
=\sum_{k>0} \left[\frac{u_k (\xi) v_k^{*}(\xi')}{z-E_k}
	+\frac{v^*_k (\xi) u_k(\xi')}{z+E_k} \right], \notag\\
G_{vu}(z;\xi,\xi') 
=\sum_{k>0} \left[\frac{v_k (\xi) u_k^{*}(\xi')}{z-E_k}
	+\frac{u^*_k (\xi) v_k(\xi')}{z+E_k} \right], \notag\\
G_{vv}(z;\xi,\xi') 
=\sum_{k>0} \left[\frac{v_k (\xi) v_k^{*}(\xi')}{z-E_k}
	+\frac{u^*_k (\xi) u_k(\xi')}{z+E_k} \right].
	\label{eq:Guv}
\end{flalign}

For the zero temperature, the densities are given by
Eq. (\ref{eq:rho-kappa}) with $f_{kk'}=f_k\delta_{kk'}=0$,
namely, $\rho_0=V^*V^T$ and $\kappa_0=V^*U^T$.
Consider a contour $C_1$ that encloses the section
$[-E_\mathrm{cut},-E_1]$ on the real axis, and $C_2$ that does
the section $[E_1,E_\mathrm{cut}]$,
where $E_1>0$ is the lowest quasiparticle energy
and $E_\mathrm{cut}$ is the cut-off energy of the pairing model space.
It is easy to find from Eq. (\ref{eq:Guv}) that 
the generalized density matrix $R_0=R_{T=0}$ can be calculated as
\begin{eqnarray}
	\frac{1}{2\pi i}\int_{C_1} G(z) \mathrm{d}z &=&
	\begin{pmatrix}
		\rho_0 & \kappa_0 \\
		-\kappa_0^* & 1-\rho_0^* 
	\end{pmatrix}
	\equiv R_{T=0} ,
	\label{eq:R_1_T=0}
	\\
	\frac{1}{2\pi i}\int_{C_2} G(z) \mathrm{d}z &=&
	\begin{pmatrix}
		1-\rho_0 & -\kappa_0 \\
		\kappa_0^* & \rho_0^* 
	\end{pmatrix}
	= I-R_{T=0} .
	\label{eq:R_2_T=0}
\end{eqnarray}
Utilizing this property, the authors in Ref.~\cite{JBRW17} proposed
a coordinate-space solver of the HFB calculation.

\subsubsection{Finite temperature and Matsubara frequencies}

The idea to extend this formulation to the finite temperature
is formally straightforward.
Using the Fermi-Dirac distribution function
$f_T(z)\equiv (1+e^{\beta z})^{-1}$,
\begin{eqnarray}
	&&\frac{1}{2\pi i}\oint_{C_1+C_2} f_T(z) G(z) \mathrm{d}z
	= R_T ,
	\label{eq:R_1} \\
	&&\frac{1}{2\pi i}\oint_{C_1+C_2} f_T(-z) G(z) \mathrm{d}z
	= I-R_T ,
	\label{eq:R_2}
\end{eqnarray}
where we assume that the contour $C_1$ ($C_2$)
is confined in the left (right) half plane of $\mathrm{Re}(z) < 0$
($\mathrm{Re}(z) > 0$).
Note that $f_T(-z)=1-f_T(z)$.
Equations (\ref{eq:R_1}) and (\ref{eq:R_2}) are formally correct, however,
it is not so useful in actual numerical calculations.
Since we do not know the lowest quasiparticle energy $E_1$, we must adopt the contours,
$C_1$ and $C_2$, passing through the origin $z=0$.
The Fermi-Dirac function $f_T(z)$ has poles at the Matsubara frequencies
$z=i\omega_n= i(2n+1) \pi /\beta$ with the integer $n\gtrless 0$.
At lower temperature ($\beta\rightarrow\infty$),
the Matsubara poles are closer to the origin, and
the numerical integration becomes more demanding.

\begin{figure}[t]
	\includegraphics[width=0.45\textwidth]{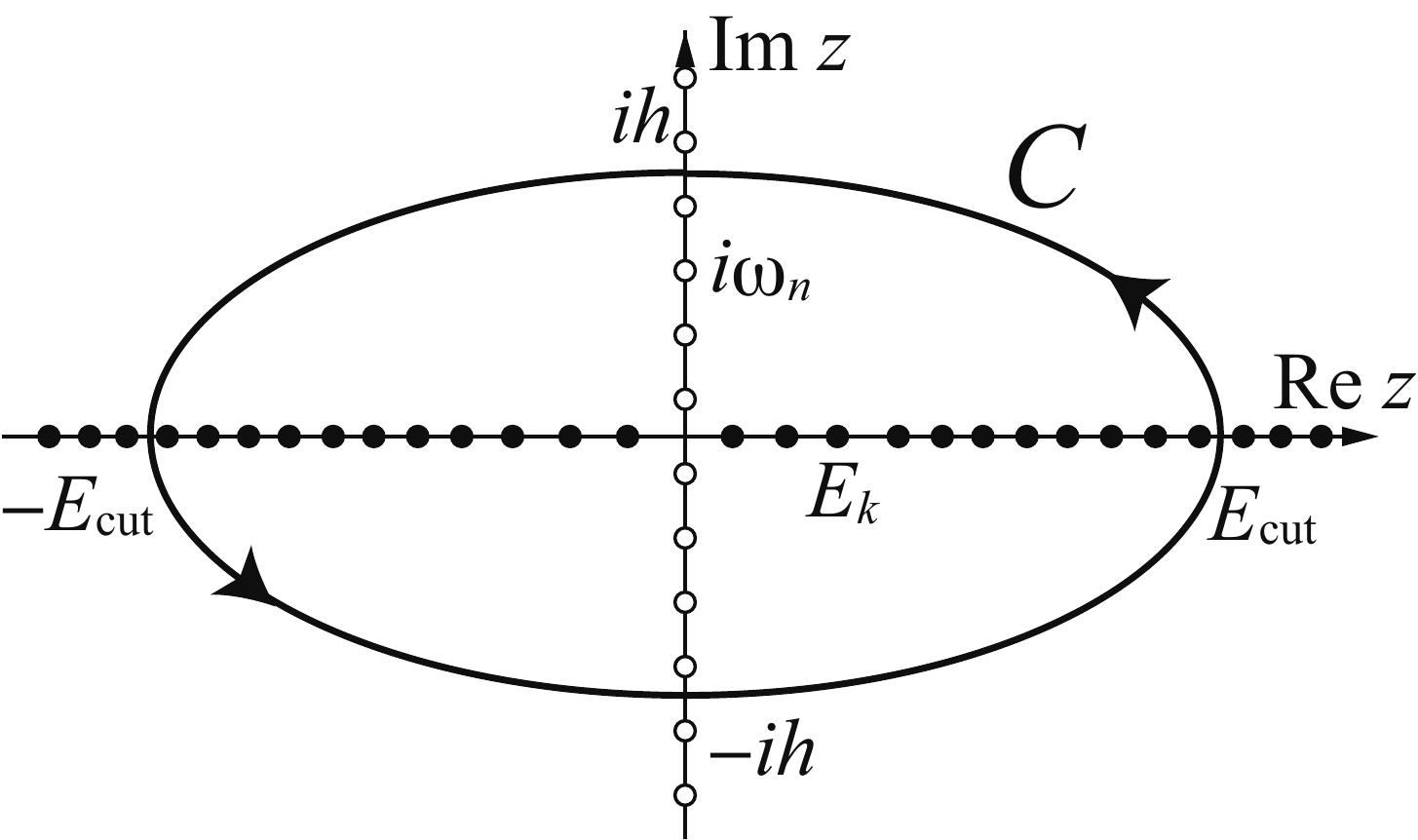}
	\caption{Schematic illustration of the contour $C$ and
	poles of the Green's function $G(z)$ (closed circles)
	and of the Fermi-Dirac function $f_T(z)$ (open circles)
	in the complex plane.}
	\label{fig:contour_1}
\end{figure}

The integrand in Eqs.~(\ref{eq:R_1}) and (\ref{eq:R_2}) is a smooth function far away from
the real axis.
Even near the Matsubara frequencies on the imaginary axis, the absolute value is reduced as $1/z$.
Therefore, numerically, the contour integration is easier with a contour further away
from the real axis.
We consider here the contour $C$ that encloses the section
$[-E_\mathrm{cut},E_\mathrm{cut}]$ on the real axis
and $[-ih,ih]$ on the imaginary axis.
See Fig.~\ref{fig:contour_1}.
Since the function $f_T(\pm z)$ has residues $\mp \beta^{-1}$ at $z=i\omega_n$,
we have
\begin{eqnarray}
	R_T &=&
	\frac{1}{2\pi i}\oint_{C} f_T(z) G(z) \mathrm{d}z
	+ \frac{1}{\beta} \sum_{|\omega_n|<h} G(i\omega_n) ,
	\label{eq:R_3} \\ 
	I-R_T &=&
	\frac{1}{2\pi i}\oint_{C} f_T(-z) G(z) \mathrm{d}z 
	- \frac{1}{\beta} \sum_{|\omega_n|<h} G(i\omega_n) .
	\label{eq:R_4}
\end{eqnarray}
The sum of these leads to an identity for the Green's function
\begin{equation}
	\frac{1}{2\pi i}\oint_{C} G(z) \mathrm{d}z = I .
\end{equation}

According to Eqs. (\ref{eq:R_3}) and (\ref{eq:R_4}),
the normal and pair densities are calculated in various ways.
Since we parallelized the computation with respect to the second argument $\xi'$,
each processor can calculate column vectors of Eq. (\ref{eq:green2x2}) with fixed $\xi'$.
For instance, from Eq. (\ref{eq:R_4}),
\begin{flalign}
\rho_T^* (\xi,\xi') = \frac{1}{2\pi i} \oint_C 
	\frac{G_{vv}(z;\xi,\xi')}{1+\exp \left(-\beta z \right)}
	\mathrm{d}z \notag \\
- k_B T\sum_{\left|\omega_n\right|<h}
	G_{vv}(i\omega_n;\xi,\xi') ,
	\label{eq:rho_G} \\ 
-\kappa_T (\xi,\xi') = \frac{1}{2\pi i} \oint_C 
	\frac{G_{uv}(z;\xi,\xi')}{1+\exp \left(-\beta z \right)}
	\mathrm{d}z \notag \\
- k_B T\sum_{\left|\omega_n\right|<h}
	G_{uv}(i\omega_n;\xi,\xi') ,
	\label{eq:kappa_G}
\end{flalign}
with
\begin{flalign}
\omega_n = \pm \pi k_B T, \pm 3\pi k_B T, \pm 5\pi k_B T \cdots .
\end{flalign}

The densities, Eqs.~(\ref{eq:rho_G}) and (\ref{eq:kappa_G}),
can be obtained from the solution of the linear equations (\ref{eq:shift}) 
without finding wave functions $(u_k(\xi), v_k(\xi))$.
We parameterize the contour $C$ as an ellipse of
\begin{flalign}
 z (\theta) = E_{\mathrm{cut}}\cos \theta + ih \sin \theta,
\label{eq:contour}
\end{flalign}
where $0 \leq \theta \leq 2 \pi$ and the height of ellipse $h$ is chosen
as the midpoint of two neighboring Matsubara frequencies,
\begin{flalign}
 h = 2m\pi/\beta,
\quad m=\mathrm{integer}.
\end{flalign}
In practice,
the contour integral is performed by dividing $C$ into four intervals,
$(0,\pi/2)$, $(\pi/2,\pi)$, $(\pi,3\pi/2)$, and $(3\pi/2,2\pi)$.
We adopt the Gauss-Legendre integration for each of these intervals.
The value of the integrand rapidly changes near the end points of these intervals,
$\phi=0, \pi/2,\pi,3\pi/2,2\pi$,
where the number of the Gauss-Legendre integral points increases.

\begin{figure}[t]
	\includegraphics[width=0.45\textwidth]{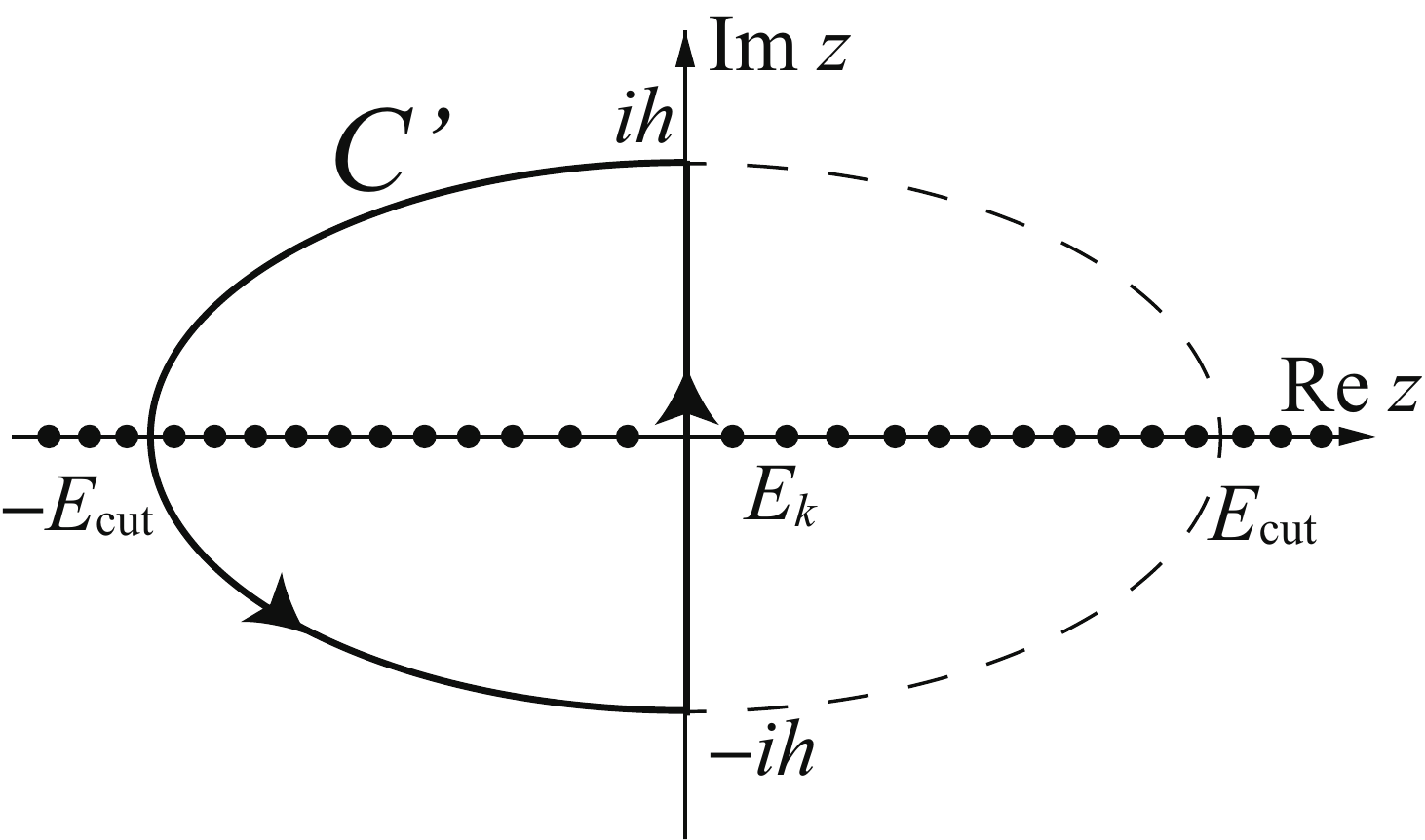}
	\caption{Schematic illustration of the contour $C'$ and
	poles of the Green's function $G(z)$ (closed circles).
	}
	\label{fig:contour_2}
\end{figure}

It is instructive to consider the limit of $T\rightarrow 0$.
In this limit, the Fermi-Dirac function $f_T(z)$ is nothing but a step function
$\theta(-\mathrm{Re}(z))$.
It vanishes in the half plane of $\mathrm{Re}(z)>0$, while
it is unity in the other half plane of $\mathrm{Re}(z)<0$.
Thus, in the first term of Eq. (\ref{eq:R_3}), the integrand can be replaced by $G(z)$,
then, the closed contour $C$ can be
changed into the open one ($\pi/2<\theta<3\pi/2$)
in the $\mathrm{Re}(z)<0$ and terminated on the imaginary axis at $z=\pm ih$.
It is known that the summation with respect to the Matsubara frequencies
becomes the integration of the Green's function on the imaginary axis at $T\rightarrow 0$.
So, the second term of Eq. (\ref{eq:R_3}) becomes
\begin{equation}
	\frac{1}{2\pi i}\sum_{|\omega_n|<h} G(i\omega_n)\frac{2\pi i}{\beta}
	\rightarrow
	\frac{1}{2\pi i}\int_{-ih}^{ih} G(z) \mathrm{d}z ,
	\quad \beta\rightarrow\infty .
\end{equation}
Therefore, Eq. (\ref{eq:R_3}) at the zero-temperature limit is identical to 
\begin{equation}
	R_T = \frac{1}{2\pi i}\int_{C'} G(z) \mathrm{d}z ,
\end{equation}
where the contour $C$ is shown in Fig.~\ref{fig:contour_2}.
In this way, we recover the zero-temperature formula, (\ref{eq:R_1_T=0}).
Following the same argument,
it is easy to obtain Eq. (\ref{eq:R_2_T=0}) from the zero-temperature limit
of Eq. (\ref{eq:R_4}).

\subsubsection{Kinetic and spin-current densities in parallel computing}

The numerical calculation is parallelized by allocating
the calculation of densities
$\rho_T(\xi,\xi')$ and $\kappa_T(\xi,\xi')$
with different $\bm{r}'$ of the second argument $\xi'=(\bm{r}',\sigma')$
on different processors.
Therefore, it is useful to eliminate the derivative $\nabla_2$
in expressions of Eqs. (\ref{eq:tau}) and (\ref{eq:J}).
The calculation of the spin-current density $\bm{J}(\bm{r})$ is done by
\begin{eqnarray}
J_x \left(\bm{r}\right) &=& - \mathrm{Im} \left[ \frac{\partial}{\partial y_1} \rho^*_T (\bm{r}\uparrow;\bm{r}\uparrow)-\frac{\partial}{\partial y_1} \rho^*_T (\bm{r}\downarrow;\bm{r}\downarrow) \right] \nonumber\\
	&+&\mathrm{Re} \left[ \frac{\partial}{\partial z_1} \rho^*_T (\bm{r}\downarrow;\bm{r}\uparrow)-\frac{\partial}{\partial z_1} \rho^*_T (\bm{r}\uparrow;\bm{r}\downarrow)  \right], \\
	J_y \left(\bm{r}\right) &=& \mathrm{Im} \left[ \frac{\partial}{\partial x_1} \rho^*_T (\bm{r}\uparrow;\bm{r}\uparrow)-\frac{\partial}{\partial x_1} \rho^*_T (\bm{r}\downarrow;\bm{r}\downarrow) \right] \nonumber\\
	&-&\mathrm{Im} \left[ \frac{\partial}{\partial z_1} \rho^*_T (\bm{r}\downarrow;\bm{r}\uparrow)+\frac{\partial}{\partial z_1} \rho^*_T (\bm{r}\uparrow;\bm{r}\downarrow)  \right], \\
	J_z \left(\bm{r}\right) &=& \mathrm{Im} \left[ \frac{\partial}{\partial y_1} \rho^*_T(\bm{r}\downarrow;\bm{r}\uparrow)+\frac{\partial}{\partial y_1} \rho^*_T (\bm{r}\uparrow;\bm{r}\downarrow) \right] \nonumber\\
	&-&\mathrm{Re} \left[ \frac{\partial}{\partial x_1} \rho^*_T (\bm{r}\downarrow;\bm{r}\uparrow)-\frac{\partial}{\partial x_1} \rho^*_T (\bm{r}\uparrow;\bm{r}\downarrow)  \right].
\label{eq:J_z}
\end{eqnarray}
The densities of Eqs.~(\ref{eq:rho}), (\ref{eq:nu}), and (\ref{eq:J})
at $\mathbf{r}=\mathbf{r}'$ can be computed by each processor
without any communication.
The kinetic density is calculated according to
\begin{flalign}
\tau (\bm{r}) = \frac{1}{2}\nabla^2  \rho (\bm{r}) 
	- \mathrm{Re}\sum_\sigma
	\nabla^2_1  \rho_T (\bm{r}\sigma;\bm{r}\sigma) .
\end{flalign}
Here, the calculation of the first time
is performed after broadcasting $\rho(\mathbf{r})$ to all
the processors.

The local densities
necessary for construction of the HFB Hamiltonian, Eqs. (\ref{eq:rho}-\ref{eq:J}),
are obtained locally ($\mathbf{r}=\mathbf{r}'$) at each processor,
then, broadcast to all the processors, to construct an updated HFB Hamiltonian.

\subsection{Shifted-COCR method}
In numerical calculations,
the most computationally demanding parts are solutions of the linear equations (\ref{eq:shift}). 
It is suitable for massively parallel computing,
because Eq. (\ref{eq:shift}) can be solved independently for
different values of $\bm{r}'$ in the calculation of $G(z;\xi,\xi')$.

Another advantageous feature of Eq. (\ref{eq:shift}) is that the shifted Krylov subspace method is
applicable to these linear equations, in which a family of the linear algebraic equations
(\ref{eq:shift}) for different values of $z$ are solved simultaneously.
For the numerical integration in Eqs. (\ref{eq:R_3}) or (\ref{eq:R_4}),
we need to solve Eq. (\ref{eq:shift}) with many values of $z$,
at discretized contour points $z_m$ $(m=0,1,...,M)$.  
In Ref.~\cite{JBRW17}, the shifted conjugate-orthogonal conjugate-gradient (COCG) method
\cite{VM90} is adopted.
In this paper, we use a similar but different algorithm,
the shifted conjugate-orthogonal conjugate-residual (COCR) method \cite{SZ11}.
The COCG method is an efficient method for positive-definite symmetric matrices.
In the present case, the Hamiltonian $H_\mathrm{HFB}$ is clearly not positive definite,
and we have found that the COCR method is more stable than the COCG method
for our purpose.
Here, we briefly present the algorithm of the shifted COCR method.

Given a symmetric matrix A, we solve the reference equation
\begin{flalign}
 A\bm{x}= \bm{b},
	\label{eq:reference}
\end{flalign}
and shifted equations
\begin{flalign}
 (A+\sigma I)\bm{x}^\sigma= \bm{b},
	\label{eq:shifted_equations}
\end{flalign}
where $\sigma$ is a complex scalar factor.
In the present case, $\sigma$ is nothing but $z$.
The reference equation (\ref{eq:reference}) is solved by COCR method.
An approximate solution $\bm{x}_{k+1}$ and its residual vector $\bm{r}_{k+1}$
in the $(k+1)$-th iteration
are calculated according to the following iterative algorithm.
\begin{flalign}
\alpha_k &= (A \bm{r}_k,\bm{r}_k)/(A \bm{p}_k,A\bm{p}_k), \\
\bm{x}_{k+1} &= \bm{x}_{k} + \alpha_k \bm{p}_{k}, \\
\bm{r}_{k+1} &= \bm{r}_{k} - \alpha_k A\bm{p}_k, \\
\beta_k &= (A \bm{r}_{k+1},\bm{r}_{k+1})/(A \bm{r}_{k},\bm{r}_{k}), \\
\bm{p}_{k+1} &= \bm{r}_{k+1} + \beta_k \bm{p}_{k}, \\
A\bm{p}_{k+1} &= A\bm{r}_{k+1} + \beta_k A\bm{p}_{k}, 
\end{flalign}
with the initial condition, $\bm{x}_0=0,\bm{r}_0=\bm{b},\alpha_0 = 1,\beta_0 = 0$.
Here, the inner product $(\bm{v},\bm{v}')$ is defined by a scalar product
$\bm{v}^T\cdot \bm{v}'$ without complex conjugation.
The matrix-vector operation is necessary only for evaluating $A\bm{r}_k$,
which is the most time-consuming part in this iteration.

We can also solve shifted equations 
	(\ref{eq:shifted_equations})
using COCR method with the same initial condition as the reference system.
However the residual vectors in shifted systems have a linear relation with the reference system,
\begin{flalign}
\bm{r}^\sigma_k = \rho^\sigma_k \bm{r}_k,
\end{flalign}
\begin{flalign}
\rho^\sigma_{k+1} = \frac{\rho^\sigma_{k} \rho^\sigma_{k-1} \alpha_{k-1}}
{\rho^\sigma_{k-1}\alpha_{k-1}\left(1+\alpha_k \sigma\right)+\alpha_k \beta_{k-1}\left(\rho^\sigma_{k-1}-\rho^\sigma_{k}\right)}.
\end{flalign}
with the initial conditions $\rho_0^\sigma=1$.
This linear relation reduces the computational cost, from $O(N^2M)$ to $O(N^2+NM)$,
where $N$ is the dimension of the matrix $A$ and $M$ is the number of complex shifts $\sigma$,
because we can avoid the time-consuming calculation of the matrix-vector product
$A \bm{r}_k^\sigma$.
The coefficients, $\alpha^\sigma_k$ and $\beta^\sigma_k$,
are also obtained from those of the reference system.
Thus, for the shifted systems, 
we simply perform the following calculations:
\begin{flalign}
\alpha^\sigma_k &= \frac{\rho^\sigma_{k+1}}{\rho^\sigma_{k}} \alpha_{k}, \\
\beta^\sigma_k &= \left(\frac{\rho^\sigma_{k+1}}{\rho^\sigma_{k}}\right)^2 \beta_{k}, \\
\bm{x}_{k+1}^{\sigma} &= \bm{x}_{k}^{\sigma} + \alpha^\sigma_k \bm{p}_{k}^{\sigma}, \label{equ:x} \\
\bm{p}_{k+1}^{\sigma} &= \bm{r}_{k}^{\sigma} - \alpha^\sigma_k \bm{p}_{k}^{\sigma}. \label{equ:p}
\end{flalign}
The iterations for the reference and the shifted systems are performed simultaneously.

In practice, it is unnecessary to calculate all the elements of
$\bm{x}_{k}^{\sigma}$ and $\bm{p}_{k}^{\sigma}$ for the shifted systems ($\sigma\neq 0$),
because physical quantities in want are often sparse in the coordinate space.
For instance, any quantity local in the coordinate requires us to calculate only
one component for each $\bm{r}'$.
This is similar to the reduced-shifted COCG method proposed in Ref.~\cite{NSFS17}.

The COCR method is designed for a symmetric matrix $A$, but
the HFB Hamiltonian is a Hermitian matrix, in general. 
Following Ref.~\cite{JBRW17},
we also transform linear Hermitian problems into real symmetric ones.
Dividing a Hermitian matrix $A$ into real and imaginary parts,
the equation $A\bm{x}=\bm{b}$ can be converted into
\begin{flalign}
 \left(
    \begin{array}{cc}
      \mathrm{Re}[A]  & -\mathrm{Im}[A] \\
      \mathrm{Im}[A] & \mathrm{Re}[A]  
    \end{array}
 \right)
 \left(
    \begin{array}{c}
	    \mathrm{Re}[\bm{x}] \\
	    \mathrm{Im}[\bm{x}] 
    \end{array}
 \right) =
 \left(
    \begin{array}{c}
	    \mathrm{Re}[\bm{b}] \\
	    \mathrm{Im}[\bm{b}] 
    \end{array}
 \right).
	\label{eq:symmetric}
\end{flalign}
Because $\mathrm{Re}A$ ($\mathrm{Im}A$) is symmetric (anti-symmetric),
the matrix in Eq. (\ref{eq:symmetric}) is a real symmetric matrix.
The shifted systems with complex scalar shifts $\sigma$ are defined as
\begin{flalign}
 \left(
    \begin{array}{cc}
      \sigma I+\mathrm{Re}[A]  & -\mathrm{Im}[A] \\
      \mathrm{Im}[A] & \sigma I+\mathrm{Re}[A]  
    \end{array}
 \right)
 \left(
    \begin{array}{c}
	    \bm{x}_1 \\
	    \bm{x}_2 
    \end{array}
 \right) =
 \left(
    \begin{array}{c}
	    \mathrm{Re}[\bm{b}] \\
	    \mathrm{Im}[\bm{b}] 
    \end{array}
 \right),
\end{flalign}
where $\bm{x}_1$ and $\bm{x}_2$ are no longer real but complex vectors.
The solution of the original problem $(\sigma I+A)\bm{x}=\bm{b}$ is 
constructed by the relation $\bm{x}=\bm{x}_1 + i\bm{x}_2$.

The performance of the shifted-COCR method will be shown in Sec.~\ref{sec:shifted-COCR}.
In practice, it is not necessary to obtain a full convergence of the shifted-COCR method
with all the complex shifts $\sigma$,
because what we need is the accurate estimation of the densities, $\rho$ and $\kappa$,
by the contour integration, Eqs.~(\ref{eq:rho_G}) and (\ref{eq:kappa_G}).
Therefore, we calculate the densities every 100 iterations, and estimate the
difference between the ``old'' and the ``new'' densities,
$\delta\rho \equiv \rho^{(\mathrm{new})} - \rho^{(\mathrm{old})}$ and
$\delta\nu \equiv \nu^{(\mathrm{new})} - \nu^{(\mathrm{old})}$.
Then, the convergence condition is set as follows:
\begin{equation}
	\left|\delta\rho(\bm{r}'\uparrow)\right| < \epsilon_1,
	\quad
	\left|\delta\nu(\bm{r}'))\right| < \epsilon_1' ,
	\label{eq:condition_1}
\end{equation}
\begin{equation}
	\left|\frac{\delta\rho(\bm{r}'\uparrow)}
	{\rho^{(\mathrm{old})}(\bm{r}'\uparrow)}\right| < \epsilon_2,
	\quad
	\left|\frac{\delta\nu(\bm{r}')}{\nu^{(\mathrm{old})}(\bm{r}')}\right|
	< \epsilon_2' ,
	\label{eq:condition_2}
\end{equation}
with $\epsilon_1=10^{-8}$ fm$^{-3}$, $\epsilon_1'=10^{-6}$ fm$^{-3}$, and
$\epsilon_2=\epsilon_2'=10^{-6}$.
We stop the COCR iteration when either Eq. (\ref{eq:condition_1}) or (\ref{eq:condition_2})
is satisfied.

\subsection{Self-consistent solutions}
\label{sec:self-consistent_solutions}

The iterative calculation is performed according to the following procedure.

\begin{enumerate}
\item \label{step:1} Input the initial densities and chemical potentials, 
$ V^{(i)} =  \{\rho^{(i)}(\bm{r}), 
		\nu^{(i)}(\bm{r}), 
		\tau^{(i)} (\bm{r}), 
		\bm{J}^{(i)}(\bm{r}), 
		\mu_q^{(i)} \} \  (i=0).
$
\item \label{step:2} Calculate the HFB Hamiltonian,
		$h_{\sigma\sigma'} (\bm{r})$ and $\Delta_q (\bm{r})$, 
		and the total energy $E^{(i)}$.

\item \label{step:3} Solve the shifted linear equations (\ref{eq:shift}) to determine the Green's functions $G(z)$.
\item \label{step:4} Calculate the contour integrals of Eqs. (\ref{eq:rho_G}) and (\ref{eq:kappa_G})
	to determine the densities and the chemical potentials, $V_\mathrm{out}^{(i)}$.
	The updated chemical potential is given by
$(\mu_q)_\mathrm{out}^{(i)} = \mu_q^{(i)}
+ \alpha_{0} \tanh \left[\alpha_{1}/\alpha_{0} \left(\left<N_q \right>-N_q\right)\right] $.
 \item \label{step:5} Calculate the energy $E_{\mathrm{out}}$.
If the convergence condition, $|E_{\mathrm{out}}-E^{(i)}_{\mathrm{in}}|<\eta$,
is satisfied, the iteration stops.
\item \label{step:6} Determine the new densities and chemical potentials using
	the modified Broyden mixing, \\
$V^{(i+1)} = V^{(i)} + \sum_{j=i-n}^i w_j [V_{\mathrm{out}}^{(j)}-V^{(j)}]$.
Go back to the step \ref{step:2}. 
\end{enumerate}
In Step.\ref{step:4}, $\alpha_{0}$ and $\alpha_{1}$ are the parameter to ensure convergence,
whose typical values are $\alpha_{0}=5$ and $\alpha_{1}=0.1 \sim 1.0$.
In Step.\ref{step:6}, $w_j$ are obtained by the modified Broyden method
\cite{BBFHNSS08}.
We store the densities and the chemical potentials of last $n$ iteration for the Broyden mixing,
with $n=10$.

\section{\label{sec:results}Numerical results}

We present a few benchmark results of the FT-HFB calculation in this paper,
with some numerical details and its computational performance.
In the following calculations, 
the Skyrme energy density functional of SLy4 \cite{CBHMS97} is adopted.
The pairing energy functional is in a form of Eq. (\ref{eq:pairing_energy})
with the bare pairing strength $g_0=-250$ MeV fm$^3$.
This pairing energy functional well reproduces the two-neutron separation energies
for Sn and Pb isotopes \cite{YB03}.
We adopt either the square or the rectangular box with the square mesh,
and impose the periodic boundary condition.

\subsection{Symmetry restriction}

We have constructed a computer program of the full 3D coordinate-space representation.
This is used for the FT-HFB calculation for $^{146}$Ba
in Sec.~\ref{sec:results_Ba}.
However, in order to speed up the calculation,
the reflection symmetry with respect to the three planes,
$x=0$, $y=0$, and $z=0$, is assumed 
in the calculations in Secs.~\ref{sec:results_Hg} and \ref{sec:results_fcc}.

In the conventional calculation of the quasiparticle wave functions,
to be benefited by this restriction, 
we need to take care of symmetry properties of the wave functions.
Each of the three types of reflection produces eigenvalues
(quantum numbers) of $\pi_k=\pm 1$ ($k=x,y,z$) for the wave functions.
Therefore, the HFB Hamiltonian is block-diagonal into eight blocks,
which means that the dimension of each block is 1/8 of the
full 3D Hamiltonian matrix.
The coordinate space can be also reduced into the first octant
$(x>0, y>0, z>0)$, however,
we need to impose a proper symmetry
on the wave functions of each block;
$(\pi_x,\pi_y,\pi_z)=(+++)$, $(++-)$, $\cdots$, $(---)$.

In contrast,
the symmetry restriction can be treated much easier in the present
Green's function method,
because we do not calculate the quasiparticle wave functions.
The Hamiltonian is invariant with respect to the three reflections.
Therefore, the Hamiltonian in the first octant
can be simply copied to the other spatial regions.
Then, Eq. (\ref{eq:shift}) is solved only for $G(z;\xi,\xi')$ with
$\bm{r}'$ in the first octant.
This reduces the computational cost into 1/8 of the full calculation.
Here, we do not need to take care of different symmetry properties
of the wave functions, according to their quantum numbers.

\subsection{Performance of shifted-COCR method}
\label{sec:shifted-COCR}

In this paper, we adopt the shifted-COCR method for the solution of
the linear algebraic equations (\ref{eq:shift}).
In contrast,
the shifted-COCG method was adopted for the zero-temperature
HFB calculation in Ref.~\cite{JBRW17}.
First, let us show differences in their convergence behavior.

\begin{figure}[t]
	\includegraphics[width=0.45\textwidth]{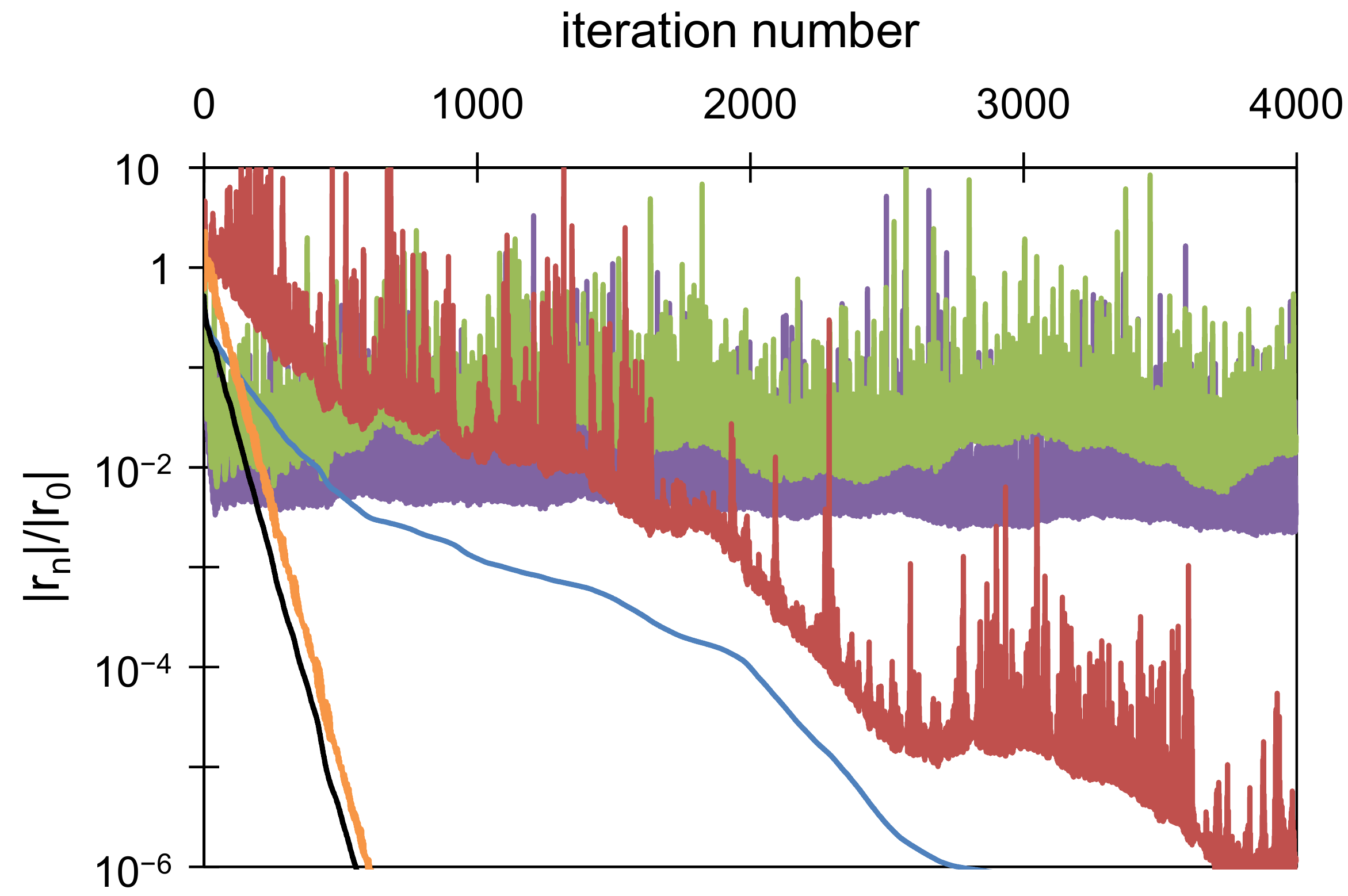}
	\caption{
The convergence behavior of the shifted COCG and shifted COCR methods
for solutions of Eq. (\ref{eq:shift}).
Three typical points, $z(\theta=0)$ (COCG/COCR: purple/green) and $z(\pi/2)$ (orange/black)
on the contour of Eq. (\ref{eq:contour})
with $E_\mathrm{cut}=100$ MeV and $h=16\pi k_B T$,
in addition to the reference point $z=0$ (red/blue),
have been taken as examples.
The norms of the residual vectors $|\bm{r}_k^\sigma/\bm{r}_0^\sigma|$
are shown as functions of the iteration number.
This is a case of the FT-HFB calculation for
the center of mass of a $^{146}$Ba nucleus ($\bm{r}'=\bm{r}_c$)
with the temperature of $k_B T=200$ keV.
The mesh and box sizes are the same as those in Sec.~\ref{sec:results_Ba}.
	}
	\label{fig:convergence}
\end{figure}

In Fig.~\ref{fig:convergence}, 
we show an example of convergence properties.
The HFB Hamiltonian $H_\mathrm{HFB}$ at the converged solution is
used for showing performance of the shifted-COCR and the shifted-COCG methods
to solve Eq. (\ref{eq:shift}).
The pure imaginary shift of $z=ih=16\pi i k_B T$ ($\theta=\pi/2$) with the COCR method shows
the fastest convergence.
The convergence behavior at $z=0$ best demonstrates superiority of
the COCR method over the COCG.
The convergence with the COCR is faster by about 1000 iterations than the COCG.
Moreover, it indicates a monotonic decrease of the residue, while the COCG shows
a strong oscillating behavior.
In general, we may expect that the convergence is slower when the shift $z$ is closer to
the pole of the integrand.
In this respect, $z=E_\mathrm{cut}$ represents the worst case.
In fact, Fig.~\ref{fig:convergence} shows that the solution at $z=E_\mathrm{cut}=100$ MeV ($\theta=0$)
fails to converge within 4000 iterations in both the COCG and the COCR methods, and the residue
$|\bm{r}_n^\sigma/\bm{r}_0^\sigma|$ keeps oscillating between $10^{-2}$ and 1.
This is not a serious problem in the calculation, because
the quasiparticle states with $E_k\approx E_\mathrm{cut}$ hardly contributes to the densities.
In addition, moving $z$ away from the real axis, the convergence property is quickly improved.

\begin{figure}[tb]
	\includegraphics[width=0.45\textwidth]{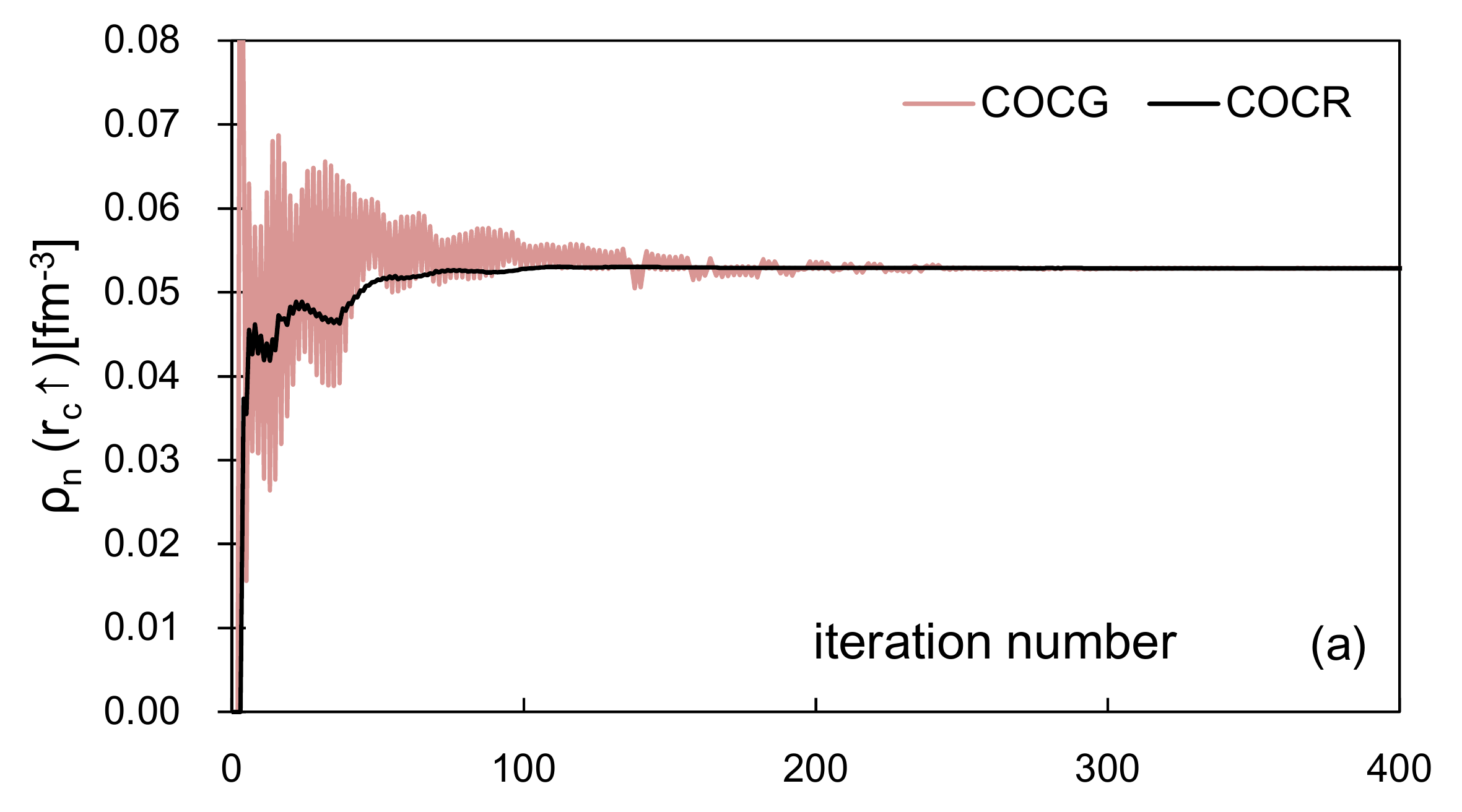}
	\newline\hspace*{-10pt}
	\includegraphics[width=0.45\textwidth]{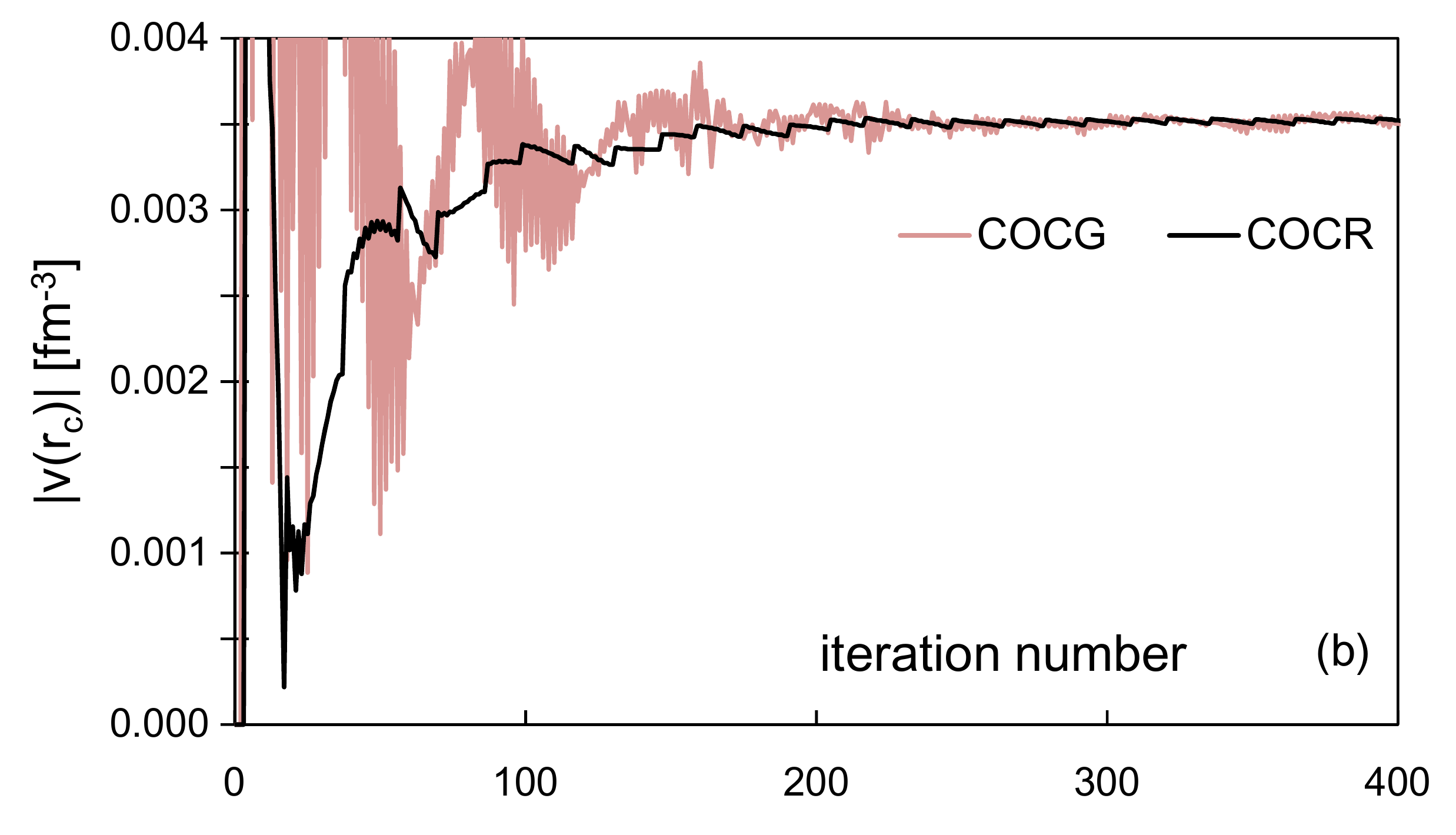}
	\caption{
		The convergence behavior of (a) the neutron spin-up density
		$\rho(\bm{r}'\uparrow,\bm{r}'\uparrow)$,
		and (b) the neutron pair density
		$\nu(\bm{r}')=\kappa(\bm{r}'\uparrow,\bm{r}'\downarrow)$
		at the center of mass
		of a $^{146}$Ba nucleus ($\bm{r}'=\bm{r}_c$) with $k_B T=200$ keV.
		The shifted COCG method is shown by red lines and the shifted COCR methods
		by black lines.
		See text for details.
	}
	\label{fig:convergence_densities}
\end{figure}

We calculate the normal and pair densities,
$\rho(\bm{r}\uparrow)$ and $\nu(\bm{r})$,
using the Green's function $G(z)$ at each iteration before the convergence.
In Fig.~\ref{fig:convergence_densities},
we show the densities as functions of the iteration number for the COCG and the COCR methods.
Note that the HFB Hamiltonian $H_\mathrm{HFB}$ is not updated during the iteration.
The densities are well converged after a few hundreds of iterations in the scale of
Fig.~\ref{fig:convergence_densities}.
Again, the convergence is faster and more stable with the COCR method than with the COCG.
Even though the shifted COCR/COCG methods fail to converge at $z$ very near $E_\mathrm{cut}$,
the densities constructed by the contour integrals, Eqs. (\ref{eq:rho_G}) and (\ref{eq:kappa_G}),
can be accurately estimated.
To achieve the convergence condition of Eq. (\ref{eq:condition_1}),
typically about 3000 iterations are required.

In order to reach the final self-consistent solution,
another self-consistent iteration is necessary.
In Sec.~\ref{sec:self-consistent_solutions},
we present the iterative procedure with the Broyden mixing to obtain
the self-consistent solutions.
This self-consistent iteration requires about several tens to hundreds of iterations.

\subsection{Octupole deformation in $^{146}$Ba at finite temperature}
\label{sec:results_Ba}

We perform the FT-HFB calculation for a neutron-rich nucleus of $^{146}$Ba as the
first benchmark calculation.
The full 3D box of the lattice size $25\times 25\times 30$
with the square mesh of $\Delta x=\Delta y=\Delta z =1$ fm
is used in the calculation.
The calculations are performed with the temperature spacing of $k_B T=100$ keV.

The nucleus of $^{146}$Ba has $Z=56$ and $N=90$,
which is in a region of strong octupole correlations \cite{BN96}.
The excitation energies of negative-parity states decrease as the neutron number
approaches to 90, and a signature of the octupole instability,
alternating parity bands, were observed in experiments at spins higher than $I=6$
\cite{Phi86,Zhu95}.
This is due to particle-hole octupole correlations associated with
$\pi [h_{11/2}(d_{5/2})^{-1}]$ and $\nu [i_{13/2} (f_{7/2})^{-1}]$.
Thus, we may expect an octupole deformed shape in the ground state of
the zero-temperature HFB theory \cite{EN17}
and it is interesting to see effects of finite temperature on its structure.

\begin{figure}[tb]
	\includegraphics[angle=90,width=0.5\textwidth]{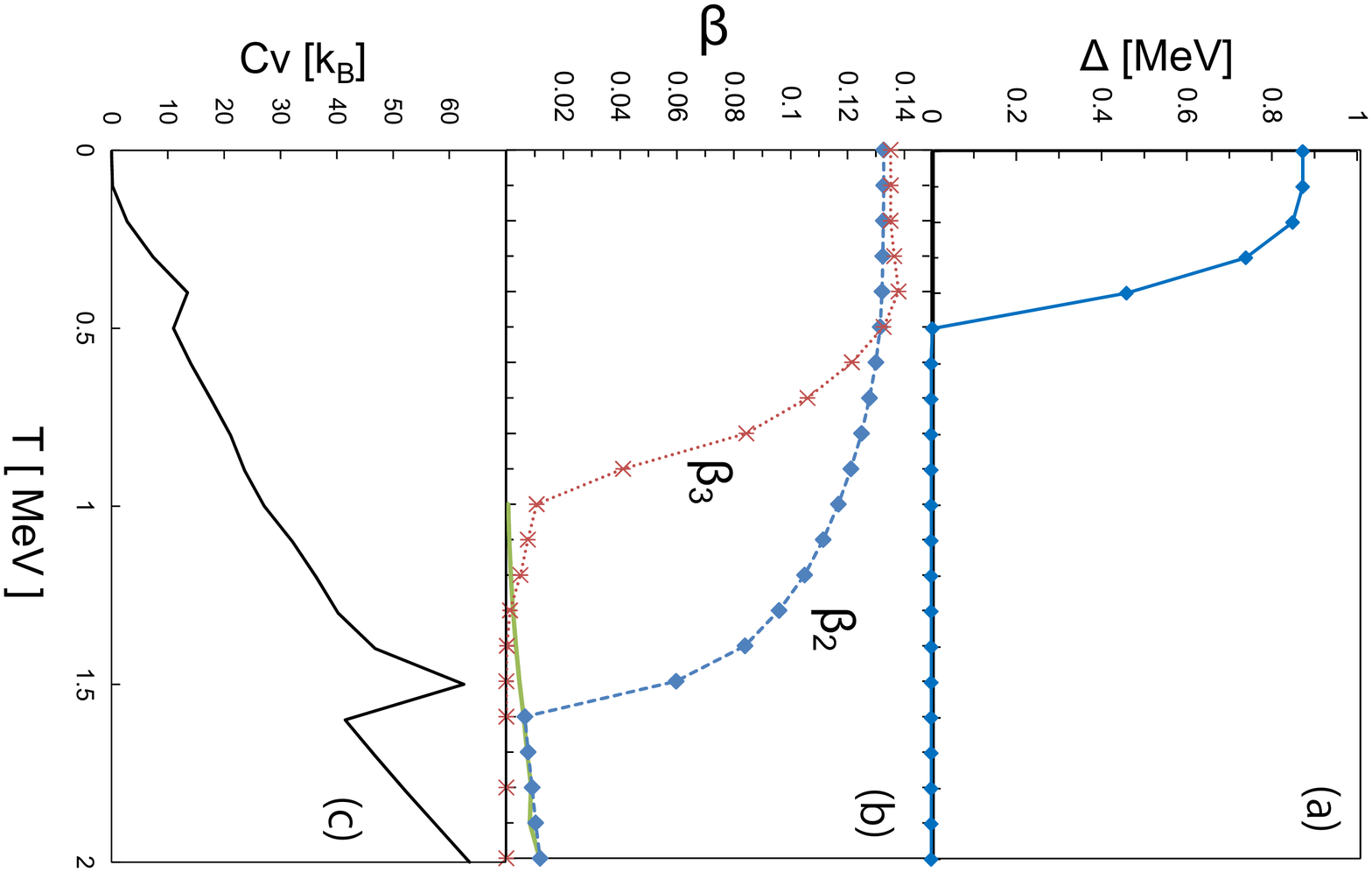}
	\caption{
		(a) Calculated neutron average paring gap,
		(b) quadrupole and octupole deformation parameters,
		and (c) specific heat as functions of temperature for $^{146}$Ba.
		In the panel (b), the quadrupole deformation of
		the dripped uniform neutrons is shown by the solid line.
	}
	\label{fig:146Ba}
\end{figure}

\begin{figure}[tb]
	\includegraphics[width=0.45\textwidth]{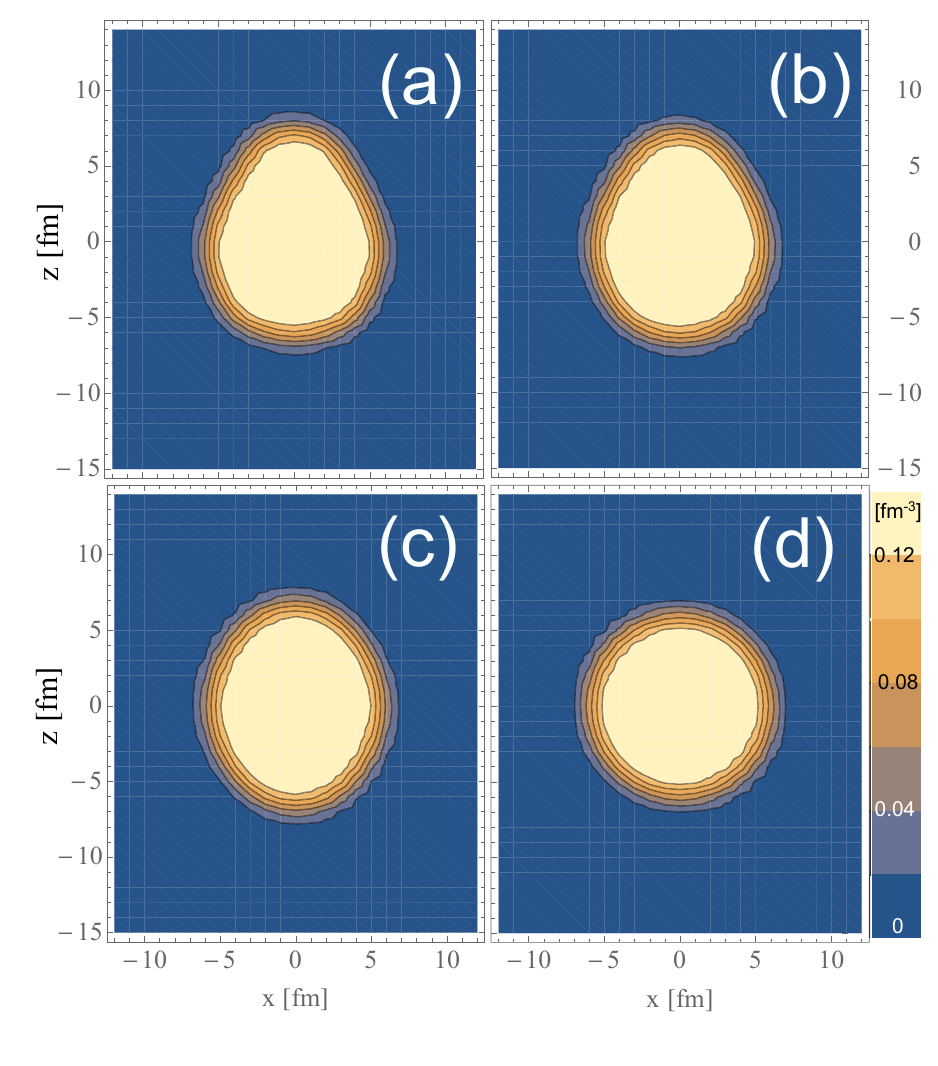}
	\caption{
	Nucleon density profiles in the $z-x$ plane for $^{146}$Ba at different
	temperature; (a) $T=0$, (b) $k_B T=0.8$ MeV, (c) $k_B T=1.2$ MeV,
	and (d) $k_B T=1.6$ MeV.
	}
	\label{fig:146Ba_profiles}
\end{figure}

The proton pair density is calculated to vanish.
We show the neutron average pairing gap in Fig.~\ref{fig:146Ba}(a).
The neutron gap is finite at low temperature but disappears at $k_B T=500$ keV.
In this calculation, the transition from super to normal phases for neutrons
is predicted at $400 <k_B T \leq 500$ keV.
In contrast, the nuclear deformation is more stable against the temperature.
At the ground state (zero temperature), the calculation predicts finite values
for both quadrupole and octupole deformations, $\beta_2\approx \beta_3\approx 0.13$.
Figure~\ref{fig:146Ba}(b) shows the temperature dependence of these deformation parameters.
At $k_B T=500$ keV, where the neutron pairing collapses, the temperature effect on
$\beta_2$ and $\beta_3$ are very little.
They are almost identical to their values at $T=0$.
Beyond $k_B T=500$ keV, the octupole deformation starts decreasing and becomes
negligibly small at $k_B T > 1$ MeV.
The quadrupole deformation is even more robust but suddenly vanishes at $k_B T=1.6$ MeV.
At temperature between 1 MeV and 1.6 MeV, the nuclear shape is almost prolate.
Beyond $k_B T=1.6$ MeV, the shape becomes spherical.
These shape changes can be clearly seen in the density distributions
in Fig~\ref{fig:146Ba_profiles}.

The quadrupole deformation is finite and slightly increases with
temperature at $k_B T>1.6$ MeV.
This is due to effects of dripped neutrons at finite temperature.
Because of the adopted rectangular box, the dripped ``free'' neutrons form the rectangular shape
which has a non-zero value of $\beta_2$.
To confirm this,
assuming a uniform density distribution of neutrons with calculated density values
at the box boundary,
we estimate the $\beta_2$ value which is shown by the solid line in 
Fig.~\ref{fig:146Ba}(b).

Finally, the specific heat $C_V(T)$ is shown in Fig.~\ref{fig:146Ba}(c).
The specific heat is estimated by the finite difference of the total energies
calculated at $k_B T\pm 0.01$ MeV.
The calculated $C_V(T)$ is approximately a linear function of the temperature $T$,
similar to that of the Fermi gas.
However, at very low temperature $T\approx 0$,
because of the proton shell gap and the neutron pairing gap, it is deviated from 
the linear dependence.
In addition, we observe sudden decreases of $C_V(T)$ at
special points of $T$, where abrupt changes in nuclear structure take place.
The first drop is associated with the collapse of the neutron pairing
at $k_B T\approx 500$ keV,
while the second one is with the shape change from prolate to spherical shapes
at $k_B T\approx 1.6$ MeV.
On the other hand, the disappearance of the octupole shape around
$k_B T\approx 1$ MeV have very little influence on it.
In contrast, the shape transition to the spherical shape
($k_B T \approx 1.6$ MeV)
leads to a significant impact on
the specific heat, a sudden decrease by more than 30 \%.
This may be due to an enhanced shell effect by
the recovered spherical symmetry.

\subsection{Shape coexistence in $^{184}$Hg at finite temperature}
\label{sec:results_Hg}

The neutron-deficient Hg isotopes are known to be a typical nuclei
showing shape coexistence phenomena \cite{Hey83,Woo92,HW11}.
Many evidences of the shape coexistence were observed, including
coexisting bands with different deformation in even isotopes and
anomalously large isotope shifts in odd-A isotopes.
Note that the Hg isotopes also exhibit superdeformed bands at high spins
\cite{JK91} on which octupole vibrations are built \cite{NMMS96}.

We have studied the temperature effect on the shape coexistence,
with the FT-HFB calculation using a constraint on the quadrupole deformation $\beta_{20}$.
The 3D box of the lattice size $30^3$
with the square mesh of $\Delta x=\Delta y=\Delta z =1$ fm
is adopted, however,
we assume the reflection symmetry with respect to the
three planes ($x=0$, $y=0$, and $z=0$), and reduce the computational cost.
The calculations are performed with different temperatures; $k_B T=0$,
0.4, 0.8, 1.6, 3.2 MeV.
The quadratic constraint on the deformation $\beta_{20}$ is
used with the spacing of $\Delta\beta_{20}=0.04$.

\begin{figure}[tb]
	\includegraphics[width=0.45\textwidth]{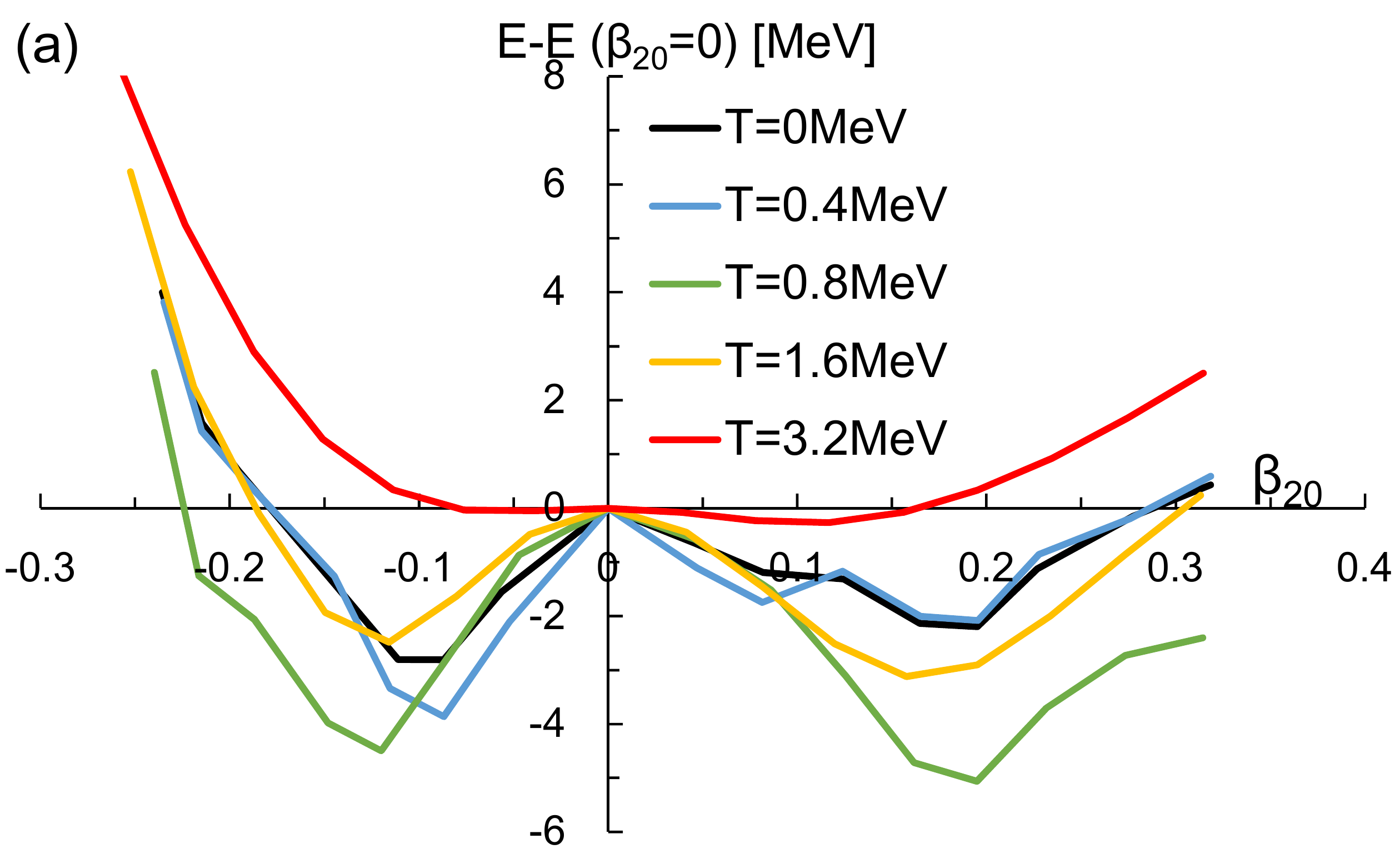}

	\vspace*{10pt}
	\includegraphics[width=0.45\textwidth]{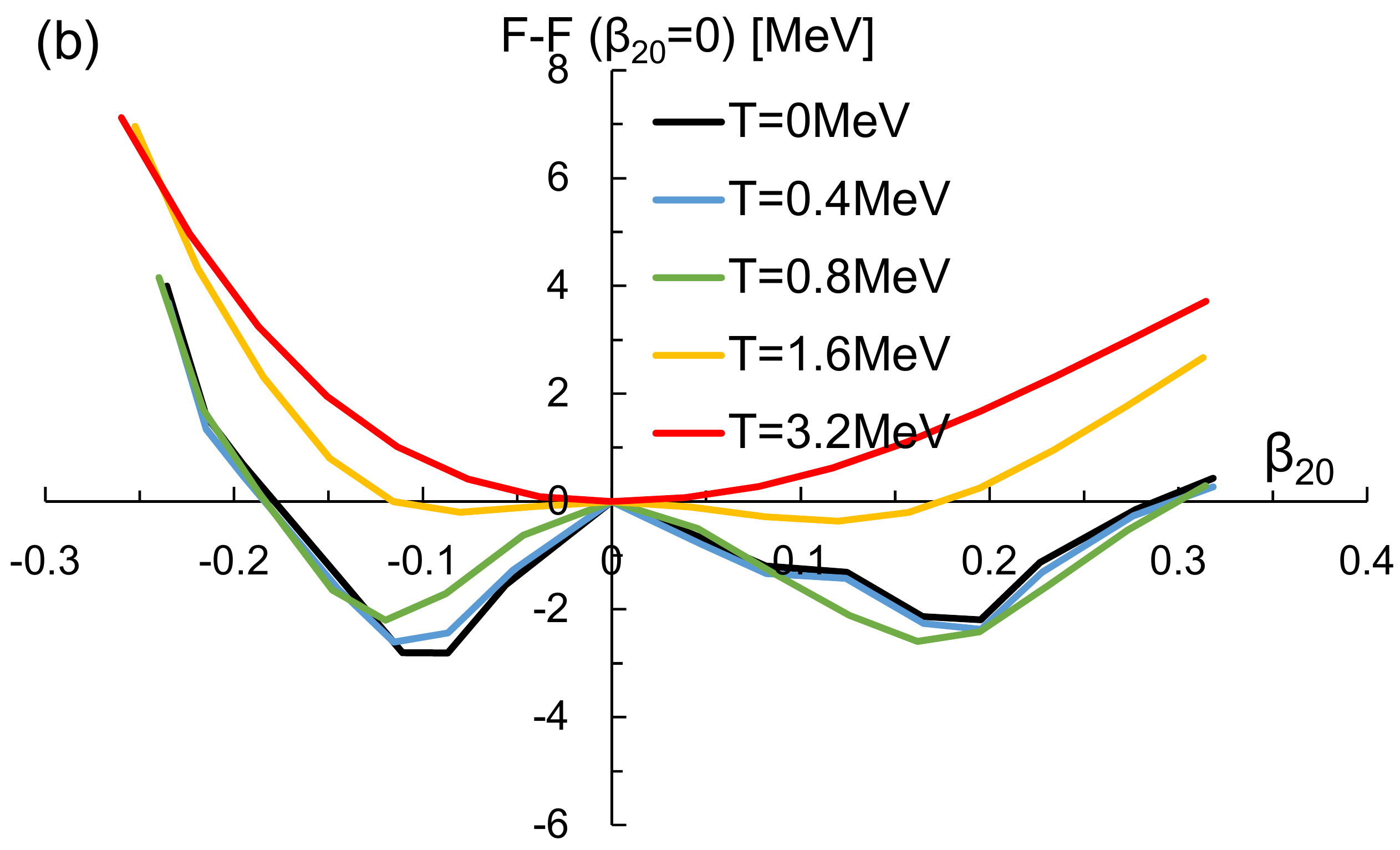}
	\caption{
		Potential energy surface, calculated with the constrained
		FT-HFB method, as a function of the quadrupole deformation,
		for $^{184}$Hg.
		The panel (a) is the total energy $E$, while the panel
		(b) is the free energy $F$.
		See text for details.
	}
	\label{fig:184Hg}
\end{figure}

Figure~\ref{fig:184Hg} shows the temperature dependence of
the potential energy surface for $^{184}$Hg.
The total energy $E(\beta,T)$ is calculated at each deformation and temperature,
then, the energy relative to the value at $\beta_{20}=0$ is plotted
in the panel (a),
while the free energy $F(\beta,T)$ is shown in the panel (b).
We can clearly see two local minima at prolate and oblate shapes.
At $T=0$, the deformation and the pairing gaps at the oblate minimum
are calculated as $\beta_{20}\approx -0.1$, $\Delta_n\approx 1.4$ MeV,
and $\Delta_p=0$.
Those at the prolate minimum
are $\beta_{20}\approx 0.2$, $\Delta_n\approx 0.8$ MeV,
and $\Delta_p\approx 1$ MeV.
The shape coexistence feature is quite robust against the finite temperature.
Although the lowest minimum at zero temperature is in the oblate side,
the prolate minima is more stable as increasing the temperature.
The oblate minima are lower than the prolate ones at $k_B T \leq 400$ keV,
while the prolate becomes lower at $k_B T \geq 800$ keV.
The main features are the same for $E(\beta,T)$ and $F(\beta,T)$,
except that a shallow prolate minimum exists for $E(\beta,T)$
but not for $F(\beta,T)$.

It is interesting to see that the non-zero temperature not necessarily favors
the spherical shape.
The potential energy surfaces $E(\beta,T)$ at $k_B T=0.8$ and 1.6 MeV indicate
deeper prolate minima than that of the zero temperature.
It is also true for the free energy $F(\beta,T)$ at $k_B T=0.8$.
This may be partially due to the pairing collapse at finite temperature.
For instance, at $k_B T=0.8$ MeV, 
the proton pairing vanishes for all the values of
deformation $\beta_{20}$.
The neutron pairing gap still has non-zero values but only
in the vicinity of the spherical shape ($\beta_{20}\approx 0$).
The vanishing pairing may lead to stronger shell energy that favors
the deformation.


We should note that the calculation of the free energy $F$
requires an additional computation.
In order to calculate $F\equiv E-TS$, we evaluate the entropy $S$ using
the formula
\begin{equation}
S=-k_B \sum_{k>0} \left[ f_k \ln f_k + (1-f_k)\ln(1-f_k) \right] ,
\end{equation}
where $f_k$ is given by Eq. (\ref{eq:occ}) in which
the quasiparticle energies $E_k$ are defined as the eigenvalues of 
the constrained HFB Hamiltonian,
namely, the mean-field approximation
to $\hat{H}-\mu\hat{N}-\lambda \hat{Q}_{20}$.
In the present Green's function method, since we do not explicitly
calculate the quasiparticle states (and their energies),
an additional diagonalization of the constrained HFB Hamiltonian
is needed after the self-consistent iteration converges.
We use the ScaLAPACK library for this diagonalization.

\subsection{Neutron-star inner crust at finite temperature}
\label{sec:results_fcc}

The neutron stars are a sort of macroscopic nucleus in the universe.
They are supposed to be synthesized by explosive stellar phenomena, such 
as supernovae.
The proto neutron stars are hot, but subsequently cooled down to
the cold neutron stars.
It would be of great interest to study neutron star matters at a variety
of temperature, especially various inhomogeneous phases predicted to exist
in the crust region near the surface.

Microscopic studies of the inner crust is theoretically very challenging,
because the calculation requires a large space in which extremely neutron-rich
nuclei and free neutrons coexist.
In addition, the energy difference between different configurations is
very small.
Thus, to predict the structure of the inner crust,
the large-scale and highly accurate calculations are needed.

In this subsection, we present our first benchmark FT-HFB calculation
for the inner crust.
The full 3D box of (45 fm)$^3$
with the square mesh of $\Delta x=\Delta y=\Delta z =1.5$ fm
is reduced by 1/8 assuming the reflection symmetry.
The calculation is performed for the temperature of $k_B T=200$ keV.
We use an initial state for the iteration with the face centered
configuration (fcc).
The adopted square box contains 4 nuclei.
We fix the neutron chemical potential as $\mu_n=10$ MeV, and
determine the proton chemical potential $\mu_p$ to satisfy the
beta equilibrium condition,
$\mu_p+m_p c^2 + \mu_e = \mu_n+m_n c^2$,
where the electrons are assumed to be uniform with the chemical potential,
$\mu_e=\sqrt{m_e c^4 + p_F^2 c^2}-e^2(3n_e/\pi)^{1/3}$.

\begin{figure}[tb]
	\includegraphics[width=0.45\textwidth]{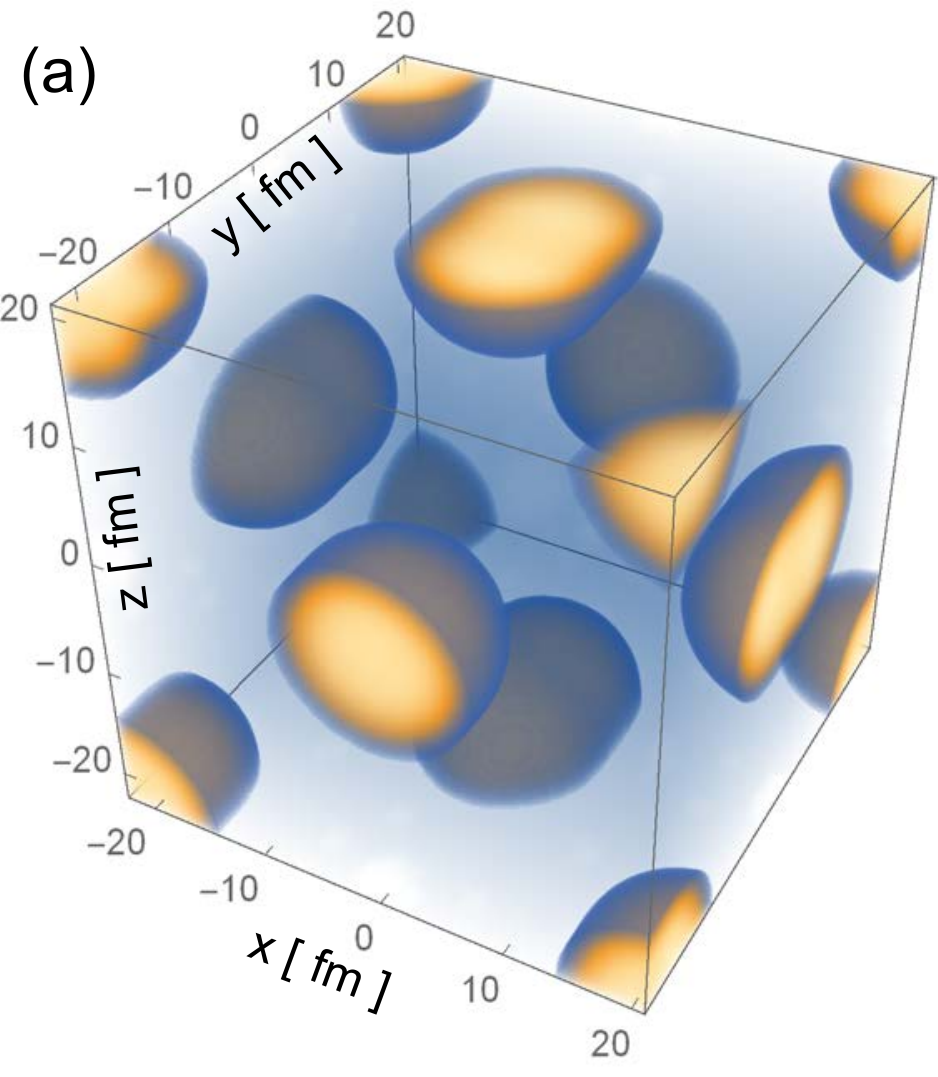}

	\includegraphics[width=0.47\textwidth]{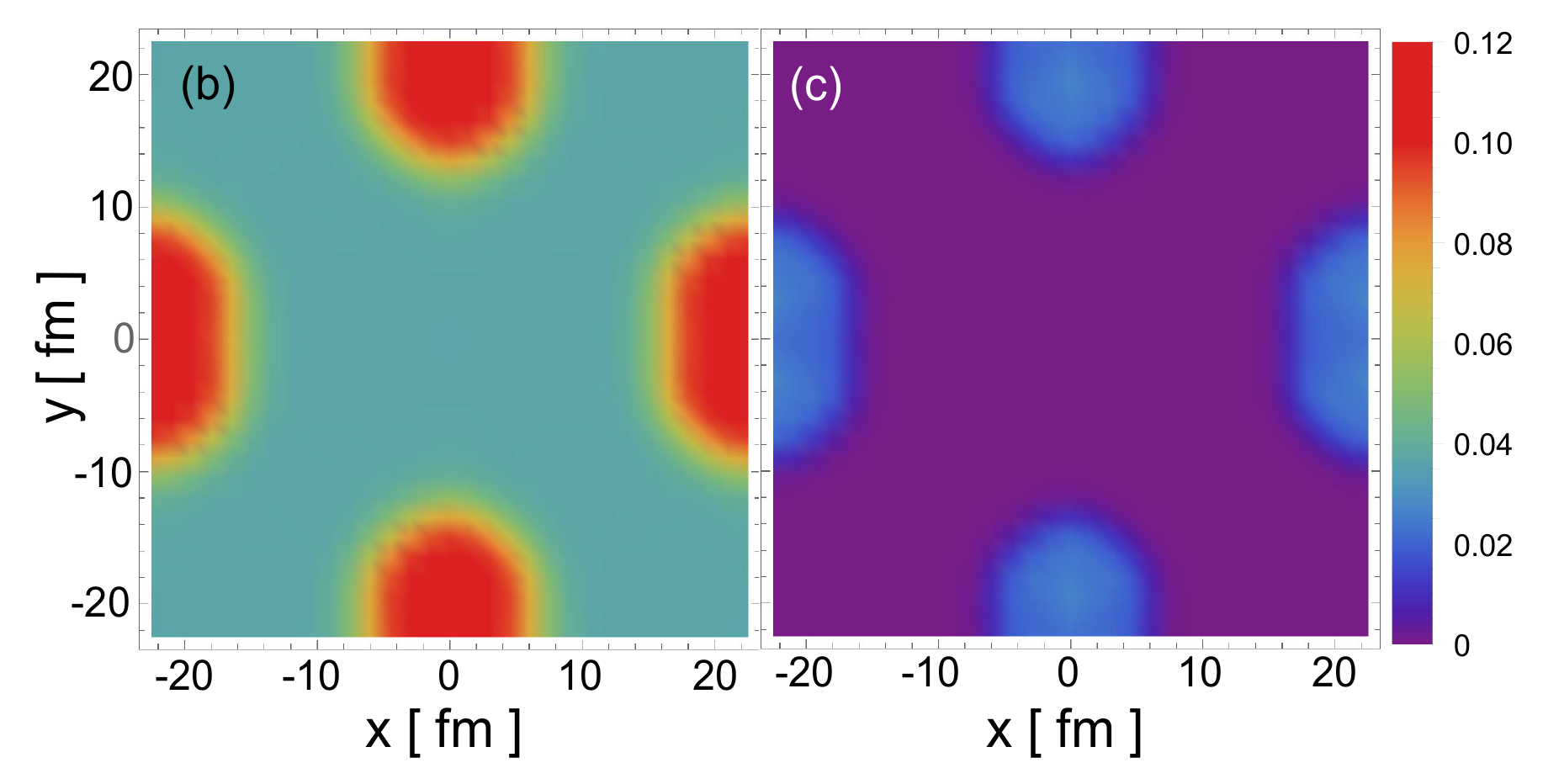}
	\caption{
		Nucleonic density distribution in the inner crust of
		neutron stars at beta equilibrium
		with the neutron chemical potential $\mu_n=10$ MeV
		and the proton chemical potential $\mu_p=-58.5$ MeV:
		(a) 3D plot of baryon density,
		(b) 2D contour plot of neutron density in the $z=0$ plane,
		and (c) the same as (b) but for protons.
	}
	\label{fig:fcc}
\end{figure}

The self-consistent procedure converges to a fcc state shown in
Fig.~\ref{fig:fcc}(a).
The neutron and proton density distributions in the $z=0$ plane are shown in
Fig.~\ref{fig:fcc}(b) and (c), respectively.
The protons are localized and form an fcc crystalline structure.
The neutrons are dripped with a large number of free neutrons.
The average nucleon density is 0.045 fm$^{-3}$,
and the lowest neutron density between the nuclei
is 0.036 fm$^{-3}$.
The obtained average pairing gap for neutrons is 1.48 MeV
and the proton gap vanishes.
The proton and neutron numbers in the box shown in Fig.~\ref{fig:fcc}(a)
are approximately 136 and 3936.
Since the protons are all confined in the crust nuclei,
we may say that emergent nuclei are very neutron-rich Se nuclei ($Z=34$).

The most interesting feature we find in this result is
that those Se nuclei are well deformed.
This can be clearly seen in the density distributions on the $z=0$ plane
shown in Fig.~\ref{fig:fcc}(b) and (c).
They are in the prolate shape.
We naively expect that, near the transition to the rod phase,
elongated nuclei may appear.
This result is a microscopic calculation to confirm this.
This is certainly a self-consistent solution of the FT-HFB
with the beta equilibrium at $k_B T=200$ keV.
However, we have not confirmed yet that this is really the optimal configuration
at the given density.
Further studies with various configurations, such as bcc and pasta phases,
is necessary to find the structure of the inner crust.

\begin{figure}[t]
	\centerline{
	\includegraphics[width=0.45\textwidth]{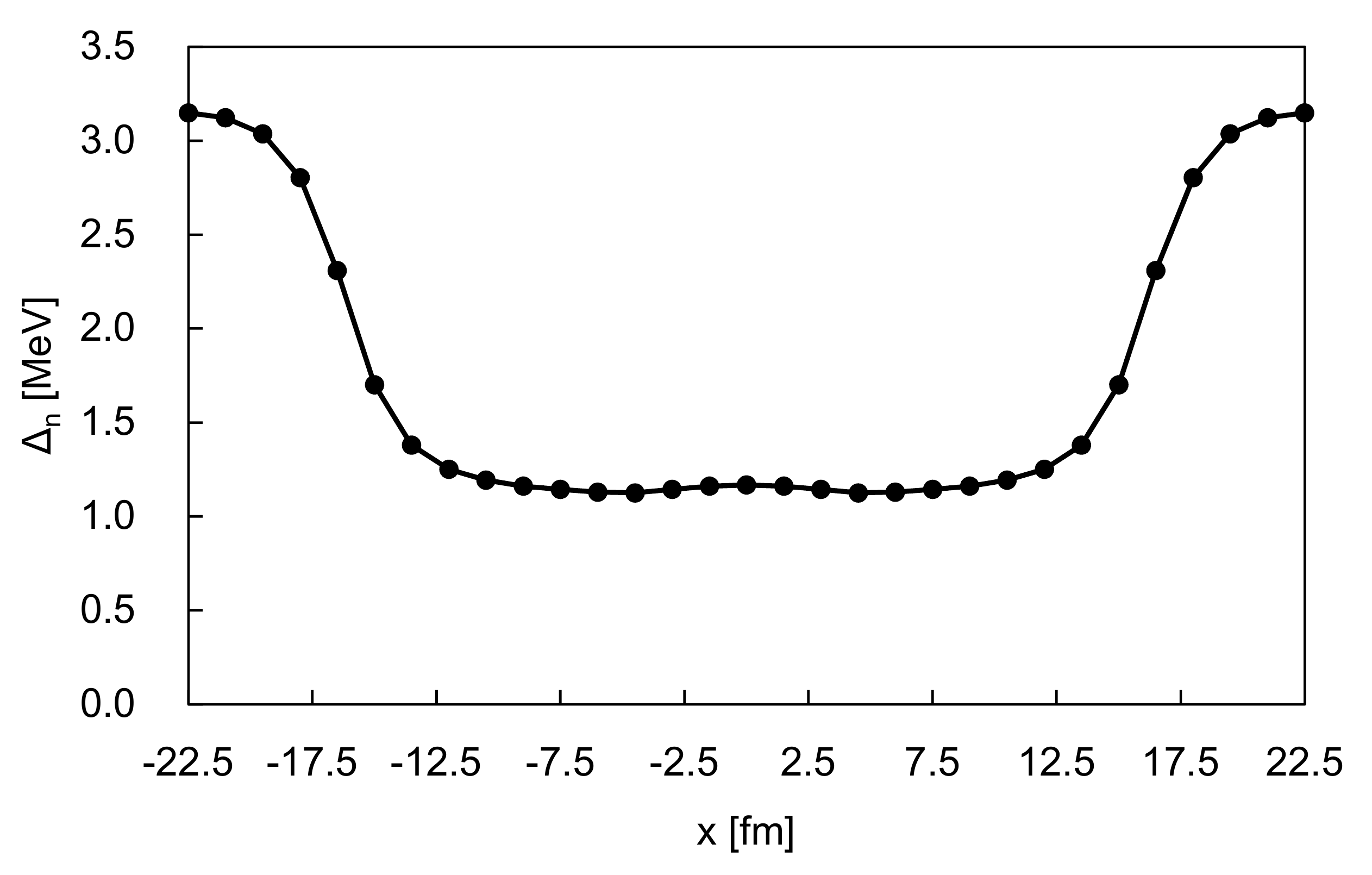}
	}
	\caption{
		Calculated neutron gap $\Delta_n(\bm{r})$ on the $x$ axis.
		The protons are clustered near the both edges.
	}
	\label{fig:Delta_n}
\end{figure}
In this state, the protons are in the normal phase, while
the neutrons are in the superfluid phase.
These superfluid neutrons are supposed to be responsible for
the pulsar glitches \cite{AI75}.
Figure~\ref{fig:Delta_n} shows the neutron pairing gap $\Delta_n(\bm{r})$
on the $x$ axis.
The left and the right ends correspond to the center of the Se nuclei,
while only dripped neutrons exist near the center ($x=0$).
The magnitude of the gap is about 1 MeV for dripped neutrons
and even larger inside the nuclei.
This is somewhat opposite to our naive expectation,
since the pairing gap calculated with the bare force for the uniform matter
is larger at low density ($\rho_n < \rho_0\approx 0.08$ fm$^{-3}$)
than at high density ($\rho_n \gtrsim \rho_0$) \cite{TT93,DHJ03}.
However, it is premature to conclude the pairing property
in the inner crust from this calculation.
Since the present pairing energy functional is
based on a simple zero-range interaction fitted to
specific regions of finite nuclei \cite{YB03},
it is desired to test the other functionals as well.
Especially, the explicit density dependence of the coupling constant $g_0$
may be necessary to simulate its density dependence \cite{Mat06,WSMBF16}.

\subsection{Summary}
\label{sec:summary}

We have developed the finite temperature Hartree-Fock-Bogoliubov (FT-HFB) method
in the three-dimensional coordinate-space representation
with the Green's function.
This is an extension of the method proposed in Ref.~\cite{JBRW17} to
the finite temperature.
In this method, neither quasiparticle wave functions nor the quasiparticle energies
are necessary to calculate.
Thus, we can avoid the diagonalization of the HFB Hamiltonian.
Various kinds of densities are evaluated by the contour integral in
the complex energy plane.
For the calculation of the Green's function with the complex energy,
we have tested two different shifted Krylov methods,
shifted COCG and COCR methods.
The shifted COCR methods are more stable and faster to reach
the convergence.

For the benchmark calculations, we showed the structure change in
$^{146}$Ba as a function of the temperature.
The octupole deformation at the ground state disappears at $k_B T\approx 500$ keV,
while the quadrupole deformation is much more stable and persists up to
$k_B T\approx 1.6$ MeV.
The effect of the shape transition to the spherical shape is clearly visible
in its specific heat.

The shape coexistence in $^{184}$Hg is also studied with the FT-HFB calculation.
It is somewhat surprising that the deformation minima become even deeper
at finite temperature compared to those at zero temperature.
The shape coexistence is quite robust with respect to
the increasing temperature and seems to sustain up to $k_B T\approx 2-3$ MeV.
The barrier height between prolate and oblate shapes is calculated
to be more than 3 MeV, even at $k_B T=1.6$ MeV.

The structure of inner crust of (hot and cold) neutron stars is a prime motivation of
the present development \cite{KN19}.
The method has a significant advantage over the conventional methods for
systems requiring such large spatial lattice sizes.
As for the benchmark, we have presented a beta-equilibrium fcc state
at the nucleon density of 0.045 fm$^{-3}$ and $k_B T=200$ keV.
Neutron-rich Se nuclei emerge and they are well deformed in the prolate shape.
The transition from spherical to deformed nuclei is an interesting issue
in the future study of the structure of the inner crust,
as a function of density and temperature.

\begin{acknowledgments}
This work is supported by JSPS KAKENHI Grant No.18H01209
and No.19H05142.
We thank Y. Futamura and T. Sakurai
for useful discussion on the shifted Krylov methods.
This research used computational resources provided by
Joint Center for Advanced High Performance Computing (JCAHPC)
through the HPCI System Research Project (Project ID: hp190031)
and through Multidisciplinary Cooperative Research Program
in Center for Computational Sciences, University of Tsukuba.
\end{acknowledgments}

\bibliography{nuclear_physics,current,myself}

\end{document}